\documentclass[%
 reprint,
 amsmath,amssymb,
 aps,
]{revtex4-2}[12pt]

\usepackage{amsmath,amssymb}
\usepackage{graphicx}
\usepackage{dcolumn}
\usepackage{bm}
\usepackage{braket}
\usepackage{physics}
\usepackage{svg}

\usepackage{breqn}

\begin{document}

\preprint{APS/123-QED}

\title{The Potential Inversion Theorem}

\author{Alec Shelley}
\email{ams01@stanford.edu}
\author{Henry Hunt}
\email{hshunt@stanford.edu}
\affiliation{Department of Applied Physics and Physics\char`,{} Stanford University\char`,{} 348 Via Pueblo\char`,{} Stanford\char`,{} CA\char`,{} 94305}


\begin{abstract}
    Quantum lattice models describe a wide array of physical systems, and are a canonical way to numerically solve the Schrodinger equation. Here we prove the potential inversion theorem, which says that wavefunction probability in these models is preserved under the sign inversion of the potential energy as long as the initial conditions occupy strictly even or odd lattice sites and are real up to a global phase. This implies that electron pairs time evolve like positronium and therefore form bound states. We simulate the dynamics of these paradoxical “electronium” pairs and show that they are bound together more strongly if their charge is increased. We show how the potential inversion theorem illustrates several seemingly unrelated physical phenomena, including Bloch oscillations, localization, particle-hole symmetry, negative potential scattering, and magnetism.
\end{abstract}

\date{August 1, 2023}

\maketitle

\section{Introduction}
Lattice hopping models are a leading tool used in condensed matter physics, atomic physics, and quantum information science to describe high-temperature superconductors \cite{gull2013superconductivity}, trapped ions \cite{blatt2012quantum}, insulators \cite{anderson1958absence}, atomic gasses \cite{kollath2007quench}, quantum computer gates, numerical solutions of the continuous Schrodinger equation, and even stomach bacteria \cite{fung2019high}. Any system of interacting particles can be described by a lattice hopping model by spatially discretizing the Schrodinger equation. This could be an approximation to a continuous process or representative of fundamentally discrete physics, like in solids. In either case, lattice hopping models are the natural tool to describe such quantum systems. Here, we present a novel insight into the physics of these models, with far-reaching implications: time evolution of probability is invariant to inversion of the potential $V(x) \rightarrow -V(x)$ given certain easily accessible initial conditions. This result provides a mechanism by which electrons can form bound “electronium” pairs with no interaction other than Coulomb repulsion. More generally, any potential can localize wavefunctions if it varies spatially by more than the allowed kinetic energy range of the lattice.

Bound electron pairs are the core ingredient for superconductivity, and in traditional BCS superconductors they form via the exchange of phonons \cite{tinkham2004introduction}. However, the exact reason why these electron pairs form in high temperature superconductors is a hotly debated topic. We find that if two electrons are placed next to each other on a lattice a distance $d$ apart, then they will remain bound for all time if $v>4Dgd$, where $D$ is the dimension of the lattice, $g$ is the hopping strength, and $v$ is the strength of their Coulomb interaction. This condition is shown to be achievable for lattice spacings typical for solids of ~5\AA. To see this effect experimentally, we expect that the best candidate materials are those with small hopping strengths and long coherence times for electrons since scattering or other decoherence breaks the electronium-positronium symmetry. 

This behavior of electron pairs on a lattice is a specific case of our potential inversion theorem. This theorem says that when initial conditions $\psi(t = 0)$ have value only on even or odd sites and are real up to a global phase, then evolution of probability is preserved under changing the sign of the potential. This can be thought of as a consequence of the Schrodinger equation conserving energy. A nearest neighbor hopping with strength $g$ has allowed kinetic energies in the range $-2g$ to $2g$, so if the potential energy varies over distance by more than $4g$, it can trap particles whether the variation is positive or negative.

This sign-independent trapping is known to occur in solid state systems in the form of Bloch oscillations \cite{leo1992observation}, where electrons cannot travel down linear potentials created by an electric field, instead oscillating back and forth. High energies halting transport is also known to occur in Anderson localization; where the high as well as low energy ``band edge" states are localized, but the intermediate energy states are delocalized \cite{abrahams201050}. These can be thought of as resonance phenomenon like Rabi flopping or damped oscillation, where too much energy can halt the process just like too little energy. Similarly, light reflects when moving from high to low index of refraction, and wavefunctions reflect when moving from high to low potential energy. All of these phenomena are illustrated by the potential inversion theorem.    

The potential inversion theorem is remarkably simple to prove, and its applications are wide-ranging. We discuss its use for lattice models, scattering, quantum field theory, spin interaction models, statistical physics, and discrete simulations of diffusive PDE's.

\section{Potential Inversion Theorem}

\label{PIT}

Consider the nearest-neighbor hopping Hamiltonian on a lattice of arbitrary dimension and size

\begin{equation}
H_+ = g\sum_{x,n}a^{\dagger}_x a_{x+n} +\sum_x V_xa^{\dagger}_x a_x \equiv T+V
\label{one}
\end{equation}

\newcommand{\comments}[1]{}

Where the sum $x,n$ runs over every lattice site and its nearest neighbors, and $a_x$ is a bosonic or fermionic annihilation operator on site $x$. For initial conditions $\psi_0$ with support only on even or odd lattice sites, $|\psi_+(x,t)|^2$ is preserved under the transformation $T \rightarrow -T$. To see this, consider $A \equiv \prod_x (-1)^{x a^{\dagger}_x a_x}$. This operator commutes with $V$ and anti-commutes with $T$. Therefore $AH_+ = H_-A$ with $H_- \equiv -T+V$. Without loss of generality, consider $\psi_0$ occupying only even sites, i.e. $A\psi_0 = \psi_0$, then we have 
\begin{equation}
    \ket{\psi_+(t)} = e^{-iH_+t}\ket{\psi_0} = e^{-iH_+t} A \ket{\psi_0}
\end{equation}
\begin{equation}
= Ae^{-iH_-t}\ket{\psi_0} \equiv A\ket{\psi_-(t)}
\end{equation}
\begin{equation}
\psi_+(x,t) = \bra{x}A\ket{\psi_-(x,t)} = (-1)^x \psi_-(x,t)
\end{equation}

And we have $|\psi_+(x,t)|^2 = |\psi_-(x,t)|^2$. If $\psi_0$ is real up to a global phase, then the Hamiltonian is time-reversible and $|\psi_+(x,t)|^2$ is preserved under the transformation $H \rightarrow -H$. Together with the $T\rightarrow -T$ symmetry, this means that probability is conserved under the transformation $V \rightarrow -V$.

This argument can be generalized to work on Hamiltonians which include longer-range hopping. Consider terms of the form $T_i = -g_i\sum_x a^{\dagger}_x a_{x+\Delta x_i}$. If $\Delta x_i$ has an even Manhattan distance, then $T_i A = A T_i$, and if $\Delta x_i$ has an odd Manhattan distance, then $T_i A = -A T_i$. This means that probability evolution will be preserved under the transformation $T_{odd} \rightarrow -T_{odd}$ which inverts the sign of all the odd-distance hopping terms. If the initial conditions are real up to a global-phase, then probability will be preserved under the transformation $T_{even} \rightarrow -T_{even}$. Specifically, any potential energy $V(x)$ diagonal in position space can be considered a zero-distance hopping.

\begin{figure}
    \includegraphics[scale = .45]{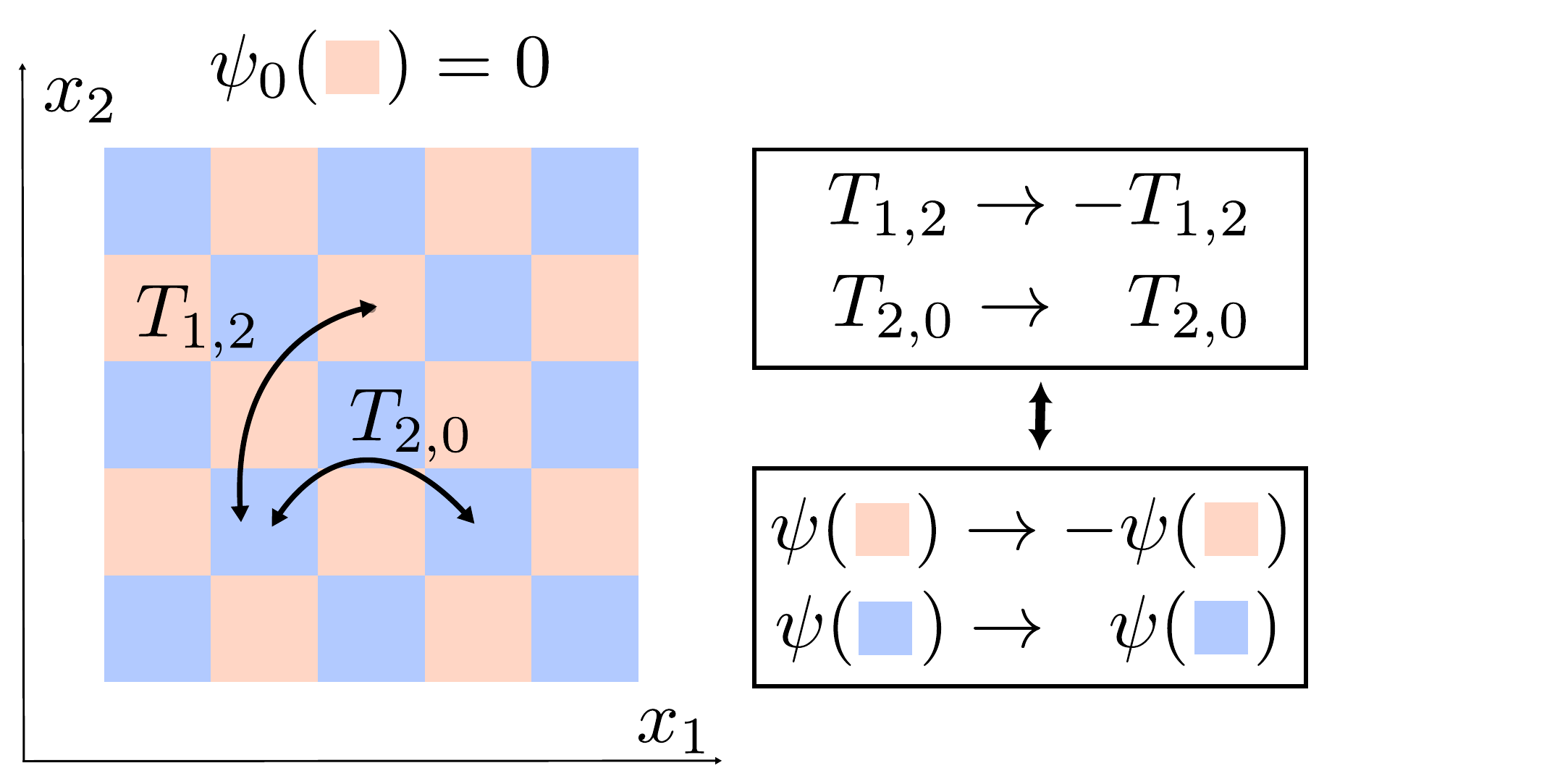}
    \caption{Potential inversion theorem for a Hamiltonian with a two-distance hopping, $T_{2,0}$ and a three-distance hopping $T_{1,2}$. If the wavefunction starts only on blue squares, then inverting the sign of the odd-distance hopping inverts the sign of the wavefunction on orange squares. If the initial conditions were real, then inverting the even-distance hopping would have the same effect except for conjugating the wavefunction everywhere.}
    \label{checkerboard}
\end{figure}

Figure \ref{checkerboard} demonstrates inversion of odd-hopping for a Hamiltonian with two- and three-distance hopping. The intuition behind the potential inversion theorem is that if the wavefunction starts on blue squares, then it takes an odd number of odd distance hoppings to get to an orange square, and an even number to get to a blue square. Therefore, inverting the sign of all odd-distance hoppings inverts the sign of the wave-function on orange squares, and preserves it on blue squares. 

Inverting odd-distance hoppings preserves energies since $H_- A\ket{E_+} = AH_+ \ket{E_+} = E_+ A\ket{E_+}$. Similarly, inverting even-distance hoppings ($H_+ \rightarrow -H_-$) inverts energies, with the highest energy eigenvectors becoming those with the most negative energies. A Hamiltonian with only odd couplings will have $HA\ket{E} = -E A \ket{E}$ and therefore have energies which come in pairs $E, -E$. This can be interpreted as a generalized particle-hole symmetry, where $A$ maps particles in the conduction band to holes in the valence band. For example, nearest neighbor hopping $H = gT$ has energies $2g \cos (k)$ on an infinite lattice, and $A$ maps $k \rightarrow k+\pi$, which takes $E_k \rightarrow -E_k$.

The potential inversion theorem allows for inversion of hoppings along specified directions. This is done by altering the $A$ operator to sense only certain dimensions in the exponent of $-1$. For example, in two dimensions, we could define $A \equiv \prod_{x,y} (-1)^{x a^{\dagger}_{x,y} a_{x,y}}$. This operator commutes with hopping terms over an even $x$ distance, and anti-commutes with hopping over an odd $x$ distance, independent of $y$ distance. Proceeding with this new $A$ as before, we have that if $\psi_0$ occupies sites with strictly even or odd x-coordinates, then probability is conserved under inverting odd x-distance hoppings. For example, this says that in a discrete setting, the Hamiltonian $-\partial_x^2-\partial_y^2 +V(x)$ has the same time evolution as $-\partial_x^2 + \partial_y^2 +V(x)$, up to a phase. This will be discussed more in a later section.

\section{Negative Potential Scattering}

\label{Neg}

The potential inversion theorem provides simple intuition about the phenomenon of negative potential scattering. Just like the counter-intuitive result that wave-functions can transmit through potential barriers, they can also reflect off of negative potential wells. Consider a wavepacket with width $\sigma$ and wavenumber $k$ coming from the left and reflecting off of a sigmoidal potential drop given by $V(x) = -\Delta E (1+\tanh{(x/\Delta x)})$ with depth $\Delta E>0$ and width $\Delta x$. As shown in \cite{garrido2011paradoxical}, necessary and sufficient conditions for a large reflection probability $R\approx 1$ are given by $1/k >> \Delta x$, $\Delta E >> k^2$, and $\sigma >> 1/k$. These conditions are exactly those in which simulating the Schrodinger equation as a lattice hopping model is likely to give accurate results.

Take a lattice spacing $a$ with $k^2 << 1/a^2 << \Delta E$, and use the finite difference formula for $\nabla^2$ to write the Schrodinger equation as $H_- = -T/a^2 + V(xa) + 1/a^2 \equiv -T+V'$. Consider the real initial conditions $\psi_0(x) \propto G_{\sigma}(x) \sin(kxa)$ with a Gaussian envelope $G_\sigma(x)$. The potential inversion theorem says 

\begin{equation}
|\bra{x} e^{-iH_-t} \ket{\psi_0}|^2 = |\bra{x} A e^{-iH_-t} \ket{\psi_0}|^2 
\end{equation}
\begin{equation}
= |\bra{x} e^{-iH_+t} A \ket{\psi_0}|^2 = |\bra{x} e^{iH_+t} A \ket{\psi_0}|^2
\end{equation}

In other words, the reflection probability is conserved under $V(x)' \rightarrow -V(x)'$ and $\psi_0(x) \rightarrow A\psi_0(x)$. Since $k << 1/a$, $A$ approximately maps $k$ to $-\pi/a$, the left edge of the Brouillon zone.  After these substitutions, the initial condition has kinetic energy $\approx (\pi/a)^2 << \Delta E$ by construction, so now it is no surprise that the wavefunction would reflect off of the potential barrier. The same logic holds for reflection off of a negative delta function potential. This also explains why light reflects off of a sharp drop in index of refraction, or why waves on a string reflect off of a drop in mass density\footnote{This is done by mapping the wave equation in the paraxial limit onto the Schrodinger equation.}. In the next section we will see how the Potential inversion theorem not only leads to negative potential scattering, but in fact negative potential \emph{confinement}, where particles can get stuck at the top of potential hills.

\comments{

Consider the discrete Schrodinger equations on L sites

\begin{equation}
    \begin{cases}
            i\partial_t \psi_{\pm}(x,t) = H_{\pm} \psi(x,t)
 & t \in (-\infty,\infty)\\
             \psi_{\pm}(x,0) = \psi_{\pm,0}(x) &  x \in \mathbb{N}, 1\leq x \leq L \\
             \psi_{+,0}(x) = (-1)^x\psi_{-,0}^*(x)
    \end{cases}
    \label{SE1}
\end{equation}

where $\psi_{\pm}(t) \in \mathbb{C}^L$ and $H_{\pm}: \mathbb{C}^L \rightarrow \mathbb{C}^L$ are $L\times L$ Hermitian matrices given by $H_{\pm} = T \pm V$. $T$ is the nearest neighbor hopping $\bra{x} T \ket {y} \equiv g\delta^y_{x+1}+g\delta^y_{x-1}$ for some $g \in \mathbb{R}$, and $V$ is any diagonal real matrix. Then $\forall t\in (-\infty,\infty)$, we have 

\begin{equation}
    |\psi_+(x,t)|^2 = |\psi_-(x,t)|^2
\end{equation}

To prove this, consider the diagonal matrix $A$ given by $ \bra{x} A \ket{y} \equiv (-1)^x \delta^y_x$. $A$ commutes with $V$ because both are diagonal, and $A$ anticommutes with $T$ since $\bra{x}\{A,T\}\ket{y}$ is

\begin{equation}
g[(-1)^x(\delta^y_{x+1}+\delta^y_{x-1})+(-1)^y(\delta^y_{x+1}+\delta^y_{x-1})] = 0
\end{equation}

This means that $AH_+ = -AH_-$, so for any eigenvector $\ket{\phi_{+,n}}$ with $H_+ \ket{\phi_{+,n}} = E_{+,n} \ket{\phi_{+,n}}$, we have $H_-(A\ket{\phi_{+,n}}) =-E_{+,n}(A\ket{\phi_{+,n}}) \equiv E_{-,n} (A\ket{\phi_{-,n}})$.

We will now expand the solution of eq. (\ref{SE1}) in terms of the complete orthonormal energy eigenbasis:

\begin{equation}
    \psi_+(x,t) = \bra{x} e^{-iH_+\hspace{.05cm}t} \ket{\psi_{+,0}}
\end{equation}

\begin{equation}
    = \sum_{n,m,x'}^L \phi_{+,n}(x) \bra{\phi_{+,n}}e^{-iH_+t} \ket{\phi_{+,m}}  \phi_{+,m}^*(x')  \psi_{+,0}(x')
    \label{contin}
\end{equation}

\begin{equation}
    =\sum_{n,x'}\hspace{.05cm} \phi_{+,n}(x)\phi^*_{+,n}(x') e^{-iE_{+,n}t} \psi_{+,0}(x')
    \label{money}
\end{equation}

Now for the time evolution under $H_-$, we get

\begin{equation}
    \psi_-(x,t)
    = (-1)^{x} \sum_{n,x'}  \phi_{+,n}(x) \phi^*_{+,n}(x') e^{iE_{+,n}t}\psi_{+,0}^*(x')
    \label{money2}
\end{equation}

The $\phi_{\pm,n}(x)$ can be taken to be real, so by comparing eq. (\ref{money}) for $\psi_+(x,t)$ to (\ref{money2}) for $\psi_-(x,t)$, we get $\psi_+(x,t) = (-1)^x \psi_-^*(x,t)$, which completes the proof.

An immediate consequence of theorem 1  is the following: if $\psi_{\pm}(x) = \psi_0(x) = p (-1)^x \psi(x)$ where $p \in\{-1,1\}$, and if $\exists \hspace{.05cm} \theta$ with $0\leq\theta<2\pi$ so that $e^{i\theta} \psi(x)$ is purely real, then the Hamiltonian is time-reversible and 

\begin{equation}
|\psi_{+}(x,t)|^2 = |\psi_{-}(x,t)|^2
\end{equation}

In other words, for wave-functions whose initial conditions have support only on even or odd sites, inverting the potential does not change time evolution of probability. Any completely localized initial condition $\ket{\psi_0} = \ket{x}$ satisfies this. 

An additional consequence of these initial conditions is $\psi_+(x,t) = \psi_-(x,-t)$
which means that inverting the potential is the same as reversing time for the wave-function. 

\comments{
\section{N-dimensional case}

In order to generalize the argument to a d-dimensional lattice, we will need to find a new operator $A$ which anti-commutes with $T$ and commutes with $V$. From there, the proof will proceed the same as in the 1-dimensional case. The d-dimensional potential inversion theorem is the following:

\begin{equation}
    |\psi_{+}(x,t)|^2 = |\psi_{-}(x,t)|^2
\end{equation}

If $\psi_\pm(x,t)$ solve the Schrodinger equations 

\begin{equation}
    \begin{cases}
            i\partial_t \psi_{\pm}(x,t) = H_{\pm} \psi(x,t)
 & t \in (-\infty,\infty)\\
             \psi_{\pm}(x,0) = \psi_{\pm,0}(x) &  x \in \mathbb{N}^d, 1\leq x_i \leq L_i \\
             \psi_{+,0}(x) = (-1)^{x_1}\psi^*_{-,0}(x) 
    \end{cases}
    \label{SE11}
\end{equation}

Where $\psi_{\pm}(t) \in \mathbb{H} \equiv \mathbb{C}^{\prod_{i = 0}^d L_i}$, where the $L_i$ are integers greater than 0. $H: \mathbb{H} \rightarrow \mathbb{H}$ is defined by $H_{\pm} \equiv T \pm V$. $V$ is defined by $\bra{x}V\ket{y} \equiv V(x) \delta^x_y$, and T is defined by 

\begin{equation}
\bra{x} T \ket{y} \equiv \bra{x_1,...,x_d} T \ket{y_1,...,y_d}
\end{equation}

\begin{equation}
    \equiv g\prod_{i = 0}^d (\delta^{y_i}_{x_i+1}+\delta^{y_i}_{x_i-1})
\end{equation}

 To prove the theorem, we use $A$ given by $\bra{x}A\ket{y} \equiv (-1)^{x_1}\delta^{x_1}_{y_1}$. In other words, $A$ is the same mapping that we used before acting on only the first component of the position $\ket{x}$. Since $A$ and $V$ are both diagonal in the position basis, they commute. We also have that $A$ anti-commutes with $T$ since 

 \begin{equation}
     \bra{x}\{A,T\}\ket{y} \equiv \bra{x}(AT+TA)\ket{y}
 \end{equation}

 \begin{equation}
     =\sum_{x'} \bra{x}A\ket{x'}\bra{x'}T\ket{y} + \bra{x}T\ket{x'}\bra{x'}A\ket{y}
 \end{equation}

 \begin{dmath}
     =(-1)^{x_1}\bra{x}T\ket{y} + (-1)^{y_1}\bra{x}T\ket{y}
 \end{dmath}

 But this is zero unless $x_1 = y_1+1$ or $x_1 = y_1-1$ due to $T$. However, in this case we have 

 \begin{equation}
     = ((-1)^{x_1}+(-1)^{x_1\pm 1})\bra{x}T\ket{y} = 0
 \end{equation}

As before, combining $[A,V] = 0$ and $\{A,T\} = 0$ gives $AH_+ = -AH_-$, so $A$ will once again provide a mapping between eigenvectors and eigenvalues of $H_\pm$ given by    $A \ket{\phi_+,n} = \ket{\phi_-,n}$ and $
    E_{+,n} = -E_{-,n}$. From here, noting once again that $A^2 = I$ and $\ket{\phi_{+,0}} = A\ket{\phi_{-,0}}$ will complete the proof the same as in the one-dimensional case.

The d-dimensional corollary is that if $\psi_{\pm,0}(x) = \psi_0(x)$ has $\psi_0(x) = p (-1)^{x_1} \psi_0(x)$ with $p \in \{-1,1\}$, and if $\psi_0(x)$ can be made a real function by multiplication with a non-zero scalar, then $|\psi_-(x,t)|^2| = |\psi_+(x,t)|^2$. In addition, we have $\psi_+(x,t) = \psi_-(x,-t)$. The extra condition of this corollary is equivalent to saying that $\ket{\psi_0}$ must be an eigenvector of $A$. Any initial conditions which have support on sites whose first component are all odd or even satisfy this. However, our choice of the using the first component of position in our definition of $A$ was arbitrary, and in fact any choice of component will work. This means that initial conditions which have support on only even or odd sites of \emph{some component} of position will satisfy the corollary. 

}

\section{General Dimension Case}
\label{gendim}

We will now generalize the potential inversion theorem to handle any number of dimensions. We start by proving the theorem for a general Hilbert space of n particles, and then show that the theorem holds for identical particles if $V$ is invariant about exchanging particles. 

The statement of the n-particle potential inversion theorem is the following:

\begin{equation}
    |\psi_+(X_1,...X_n,t)|^2 = |\psi_-(X_1,...,X_n,t)|^2
\end{equation}

If $\psi_{\pm}(X_1,...,X_n,t)$ solve the many-body Schrodinger equations

\begin{equation}
    \begin{cases}
            i\partial_t \psi_{\pm}(X_1,...,X_n,t) = H_{\pm} \psi(X_1,...,X_n,t)
\\
             \psi_{\pm}(X_1,...,X_n,0) = \psi_{\pm,0}(X_1,...,X_n) &   \\
             \psi_{+,0}(X_1,...,X_n) = (-1)^{\sum^n_{i=1}X_{i_1}}\psi^*_{-,0}(X_1,...,X_n) 
    \end{cases}
    \label{SE11}
\end{equation}

Where $t \in \{-\infty,\infty\}$, and $X_i \equiv x_{i_1},...,x_{i_d}$ is the d-dimensional position of the i'th particle. Each dimension has a maximum length $0 \leq x_{i_j} \leq L_j$ that each of the particles must be contained in. $\psi_{\pm}(t) \in \mathbb{H} \equiv \mathbb{C}^{n\prod_{i = 0}^d L_i}$, where the $L_i$ and n are integers greater than 0. $H: \mathbb{H} \rightarrow \mathbb{H}$ is defined by $H_{\pm} = T \pm V$.$\hspace{.05cm}V$ is defined by 

\begin{equation}
\bra{X_1,...,X_n} V \ket {Y_1,...,Y_n} \equiv \delta^{X_1,...,X_n}_{Y_1,...,Y_n} V(X_1,...,X_n)
\end{equation}

and $T$ is defined by $T \equiv \sum^n_{i = 1} T_i$ where 

\begin{equation}
 T_i = g(\bigotimes_{j = 1}^{i-1} I) \otimes T \otimes (\bigotimes_{j = i+1}^{n}I)
\end{equation}

$T$ can be thought of as a sum of hopping terms $T_i$ which act only on the Hilbert space consisting of the i'th particle. This is analogous to the standard definition of a kinetic energy operator for a many-body Hamiltonian. 

To prove the many-body potential inversion theorem, we again need to find an operator $A$ which anti-commutes with $T$ and commutes with $V$. This operator is given by 

\begin{equation}
    \bra{X_1,...,X_n} A \ket{Y_1,...,Y_n} \equiv (-1)^{\sum^n_{i=1}\sum^{L_j}_{j=0} x_{i_j}} \delta^{X_1,...,X_N}_{Y_1,...Y_N}
\end{equation}

$A$ can be viewed as a product of single-particle operators $A_i$ each acting on only the i'th particle. $A$ is diagonal, so it commutes with $V$, and $A$ anti-commutes with $T$ since 

\begin{equation}
 \{AT+TA\} = \sum_{i = 1}^n \{AT_i+T_iA\}
\end{equation}

And for each $i$ we have 

\begin{equation}
    \bra{X_1,...,X_n} (AT_i+T_iA) \ket{Y_1,...,Y_n}
\end{equation}

Defining $\ket{X_1,...,X_n} \equiv \ket{X}$, $\ket{Y_1,...,Y_n} \equiv \ket{Y}$, we have 

\begin{dmath}
 = \sum_{X_1',...,X_n'} \bra{X} A \ket{X_1',...,X_n'}
    \bra{X_1',...,X_n'} T_i \ket{Y}
    +\bra{X} T_i \ket{X_1',...,X_n'} \bra{X_1',...,X_n'}
    A\ket{Y}
\end{dmath}

\begin{dmath}
= (-1)^{\sum^n_{k = 1} \sum^{L_k}_{j = 0} x_{k_j}} \bra{X} T_i \ket{Y}
+(-1)^{\sum^n_{k = 1} \sum^{L_k}_{j = 0} y_{k_j}} \bra{X} T_i \ket{Y}
\end{dmath}

\begin{equation}
= (-1)^{\sum_{k,j} x_{k_j}} \bra{X} T_i \ket{Y}((-1)^{x_{i_j}}+(-1)^{y_{i_j}})
\end{equation}

The $T$ term means that the above expression is $0$ unless $x_{i_l} = y_{i_l}+1$ or $x_{i_l} = y_{i_l}-1$ for some $l$. However, if this is the case then we have 

\begin{equation}
((-1)^{x_{i_l}}+(-1)^{x_{i_l}\pm 1}) = 0
\end{equation}

So we have $\{A,T\} = 0$. This means that $A$ will once again give the mapping between $H_+$ and $H_-$ eigenstates, and the proof proceeds as before.

As a corollary, if $\ket{\psi_{\pm,0}} = \ket{\psi_0}$ is an eigenvector of $A$ and is real up to a global phase, then $|\psi_{+}(X,t)|^2 = |\psi_{-}(X,t)|^2$ and $\psi_{+}(X,t) = \psi_{-}(X,-t)$. This can be seen by completing the same steps that proved the one-dimensional corollary. 

We would now like to get a version of the potential inversion theorem valid for bosons and fermions. To do this, we will project the proof for the general many-body case onto the symmetric and anti-symmetric vector subspaces of $\mathbb{H}$ corresponding to bosons and fermions, respectively.

In order for the proof to work in these subspaces, we have to show that $A$ does not mix symmetric and anti-symmetric wavefunctions. We do this by showing that $A$ commutes with the exchange of particles. 

\begin{equation}
A \ket{X_1,...,X_n} = (-1)^{\sum^n_{i=1}\sum^{L_j}_{j=0} x_{i_j}} \ket{X_1,...,X_n}
\end{equation}

so 
\begin{equation}
    A P_{ij} = P_{ij} A
\end{equation}

Where $P_{ij}$ exchanges particles i and j.

Since symmetric and anti-symmetric wavefunctions are vector subspaces of $\mathbb{H}$, we have still have $[A,V] = 0$ and $\{A,T\} = 0$ within each space. Therefore the proof continues in each case as before except with the additional constraint that $V$ must be symmetric about the exchange of particles in order to not mix symmetric and anti-symmetric states. The corollary for the identical particle case is identical to the general n-particle case.



\section{Parity Inversion Theorem}
\label{parity}

The potential inversion theorem for nearest neighbor couplings is a consequence of the more general parity inversion theorem, which says that time evolution of eigenvectors of $A$ are preserved under the mapping which adds a minus sign to every coupling over an odd distance. We state the parity inversion theorem in $d$ dimensions with n particles as follows:

\begin{equation}
    |\psi_+(X,t)|^2 =  |\psi_-(X,t)|^2
\end{equation}

If $\psi_{\pm}$ solve the many-body Schrodinger equations

\begin{equation}
    \begin{cases}
            i\partial_t \psi_{\pm}(X,t) = H^{\pm} \psi(X,t)
\\
             \psi_{\pm}(X,0) = \psi_{0}(X) &   \\
             A \psi_{+,0}(X,t) = \psi_{-,0}(X,t) 
    \end{cases}
    \label{SE33}
\end{equation}

With the same operators and notation as before except now our Hamiltonian is made up of general couplings of the form 

\begin{equation}
    T_i = g_i \bigotimes_{j = 1}^d T_{\Delta x_j} + g_i^* \bigotimes_{j = 1}^d T_{-\Delta x_j}
    \label{hoppor}
\end{equation}

Where $T_{\Delta x_j}$ couples states seperated by $\Delta x_j$, i.e.

\begin{equation}
    \bra{x} T_{\Delta x_j} \ket{y} = \delta^y_{x+\Delta x_j}
\end{equation}

We have used to $H^{-}$ to mean the Hamiltonian whose couplings across odd distances have flipped signs (this is different than before where $H_-$ had an inverted potential).

\begin{equation}
    H^+ \equiv \Sigma_i T_i 
\end{equation}
\begin{equation}
    H^- \equiv \Sigma_i T_i (-1)^{\Sigma_{j=0}^d \Delta x_j}
\end{equation}

Note that the factor $(-1)^{\Sigma_{j=0}^d \Delta x_j}$ is $1$ for \emph{even} parity couplings, and $-1$ for \emph{odd} parity couplings. Also note that potentials $V$ are written as couplings $T_i$ of distance $0$. 

We prove the theorem by showing that $A$ commutes with even parity couplings, and anti-commutes with odd parity couplings. Consider a single coupling $T_i$ in the Hamiltonian. We have

\begin{equation}
    \bra{X} AT_i \ket{Y} 
\end{equation}

\begin{equation}
    = (-1)^{\sum^n_{i=1}\sum^{L_j}_{j=0} x_{i_j}} \bra{X}T_i \ket{Y}
\end{equation}

\begin{equation}
    =(-1)^{\Sigma_{j=0}^d \Delta x_j}  \bra{X} T_i A\ket{Y} 
    \label{myeqn}
\end{equation}

Where once again the factor $(-1)^{\Sigma_{j=0}^d \Delta x_j}$ carries the sign of the parity of $T_i$. This lets us write

\begin{equation}
    AH^+ = H^-A
\end{equation}

So we have

\begin{align}
    \psi_+(X,t) = \bra{X} e^{-iH^+ t} \ket{\psi_{+,0}} \\
    = \text{sgn}(X) \bra{X}A e^{-iH^+ t} \ket{\psi_{+,0}}  \\
    = \text{sgn}(X) \bra{X}  e^{-iH^-t} A\ket{\psi_{+,0}} \\
    = \text{sgn}(X) \bra{X} e^{-iH^- t} \ket{\psi_{-,0}} \\
    = \text{sgn}(X) \psi_-(X,t) \\
\end{align}

Where $\text{sgn}(X) \equiv (-1)^{\sum^n_{i=1}\sum^{L_j}_{j=0} x_{i_j}}$ is the parity of the position $X$. We obtain the corollary by requiring that initial conditions have support only on sites of a given parity. We now have the powerful result that the potential inversion theorem survives adding couplings of any odd distance.

In addition, by considering tensor products of the $A$ operator acting on a single dimension with identity operators, it is easy to show that the potential inversion theorem holds if odd couplings in a \emph{single} spatial dimension are inverted, as opposed to odd couplings across all dimensions. This is seen as follows: Consider the operator $A_k$ acting only on the $k$'th spatial dimension

\begin{equation}
 A_k \equiv (\bigotimes_{j = 1}^{k-1} I) \otimes A \otimes (\bigotimes_{j = k+1}^{n}I)
\end{equation}

And $T_i$, a general hopping term for particle $i$ defined by eq. (\ref{hoppor}) which couples states in the $k'th$ spatial dimension separated by distance $\Delta x_k$.

\begin{equation}
    \bra{X} \{A_k,T_i\} \ket{Y} = ((-1)^{x_{i_k}}+(-1)^{y_{i_k}}) \delta^{y_{i_k}}_{x_{i_k} \pm{} \Delta x_k}
\end{equation}

which gives

\begin{equation}
    H^{k,+}A_k = A_kH^{k,-}
\end{equation}

Where $H^{k,-}$ is defined as $H^{k,+}$ with all odd-couplings across spatial dimension $k$ flipped. In this case, the required initial conditions for $|\psi_{+}(x,t)|^2 = |\psi_-(x,t)|^2$ are 

\begin{equation}
A_k \psi_{+,0}(X,t) = \psi_{-,0}(X,t)
\end{equation} 

}

\section{Electron Pair Bound States}
\label{epbs}

A striking consequence of the potential inversion theorem is that pairs of electrons can form bound states on a lattice with no force other than Coulomb repulsion. Consider initial conditions where both electrons are localized and have the same phase. By the potential inversion theorem, the electrons evolve in time as if one of the electrons were a positron. In addition, the high energy eigenstates of the two electron Hamiltonian will resemble the low energy eigenstates of the positronium Hamiltonian, meaning that at high energies Coulomb Hamiltonians give rise to ``electronium" atomic states. Take the Coulomb Hamiltonian in D-dimensions with nearest neighbor hopping
\begin{equation}
\begin{split}
H_+ = & g\sum_{x,n,\sigma}a^{\dagger}_{x,\sigma} a_{x+n,\sigma} \\ +&\sum_{x_1,\sigma_1,x_2,\sigma_2} V_{x_1,x_2} n_{x_1,\sigma_1} n_{x_2,\sigma_2}
\label{Coulomb}
\end{split}
\end{equation}
Where the sum runs over every site $x$ and spin $\sigma$, $n_{x,\sigma} \equiv a^{\dagger}_{x,\sigma}a_{x,\sigma}$, and $V_{x_1,x_2} \equiv v/|x_1-x_2|$ when $x_1 \neq x_2$, and $v_{ons}$ otherwise.

\comments{
\begin{equation}
    \bra{X_1,X_2} T \ket{Y_1,Y_2} = g \sum_{d = 1}^D  \delta^{x_{1_d}}_{y_{1_d} \pm 1} + \delta^{x_{2_d}}_{y_{2_d} \pm 1}
\end{equation}

\begin{equation}
    = g \sum^D_{d = 1} \delta^{x_{1_d}+x_{2_d}}_{y_{1_d}+y_{2_d}\pm 1} \delta^{x_{1_d}-x_{2_d}}_{y_{1_d}-y_{2_d}\pm 1} \equiv g \sum_{d=1}^D \delta^{x_c}_{y_c \pm 1} \delta^{x_r}_{y_r \pm 1}
\end{equation}

\begin{equation}
    \equiv \bra{X_{c},X_r} T_{cm} \ket{X_{c}, X_r}
    \label{cmbasis}
\end{equation}

Where $X_c \equiv X_1+X_2$, $X_r \equiv X_1-X_2$. We can write a general two-particle wavefunction $\psi(X_1,X_2)$ in center of mass coordinates as the Fourier transform over the common coordinate:

\begin{equation}
    \psi(X_1,X_2) = \sum_{k} a_{k} e^{i k X_c} \phi_k(X_r)
\end{equation}

Here $k X_c$ is to be understood as $\vec{k} \cdot \vec{X_c}$. In this basis, we can use eq. (\ref{cmbasis}) to write the Hamiltonian as 

\begin{equation}
    \bra{X_c,X_r} H \ket{\psi_{k}}
\end{equation}
\begin{equation}
    = e^{i k X_c} [(\sum^D_{j=1} 2g\cos(k_j)T_j \hspace{.05cm} \phi_{k}(X_r)) + V(X_r) \phi_k(X_r)]
    \label{hamsep}
\end{equation}

Where $T_j \phi_k(X_r) \equiv \phi_k(X_r + 1_j) + \phi_k(X_r-1_j)$ is a sum of $\phi_k$ evaluated at the two nearest neighbors of $X_r$ in the $j$'th dimension. We see that moving to center of mass coordinates has allowed us to write the Hamiltonian as a one particle nearest neighbor hopping Hamiltonian with Coulomb potential $V$ and nearest neighbor hopping strengths $2 \cos(k_j)$. Immediately we see that for $k_j = \pm \pi/2$ there is no hopping term, and the Hamiltonian becomes diagonal in the relative position basis for the $j$'th dimension. 

 We can now investigate the case of two electrons placed next to each-other on a 1-D lattice, unable to escape off to infinity.  We start with localized initial conditions

 \begin{equation}
 \psi_0(x_1,x_2) = \delta(x_1-x_{0})\delta(x_2+x_{0})
 \end{equation}

 \begin{equation}
     = \delta(x_c)\delta(x_r-2x_0)
 \end{equation}

 Fourier transforming in the common coordinate and solving for the time evolution from the Hamiltonain in eq. (\ref{hamsep}) gives for $\psi(x_c,x_r,t)$


  \begin{equation}
     \frac{2}{\pi} \int^{\frac{\pi}{2}}_0 dk \hspace{.05cm} e^{i \frac{\pi}{2}(t+2)x_c}\cos{((k(t+1)-\frac{\pi}{2})x_c)}{\phi}_{k+\frac{\pi}{2}}(x_r,t)
 \end{equation}

 \begin{equation}
     \psi(0,x_r,t) = \frac{2}{\pi} \int^{\frac{\pi}{2}}_0 dk \hspace{.05cm} {\phi}_{k+\frac{\pi}{2}}(x_r,t)
     \label{summor} 
 \end{equation}

 Where 
 $\phi_{k+\frac{\pi}{2}}(x_r,t)$ is the solution of the 1-D nearest neighbor hopping Coulomb Hamiltonian with a hopping strength of $-2g\sin(k)$ and initial conditions localized at $x_{r,0}$. In deriving eq. (\ref{summor}), we used the parity inversion theorem to give $\phi_{k + \frac{\pi}{2}}(x_r,t) = (-1)^{x_r} \phi_{-k + \frac{\pi}{2}}$, and we approximated the discrete Fourier transform in the common coordinate as continuous.

}

\begin{figure}
    \includegraphics[scale = .565]{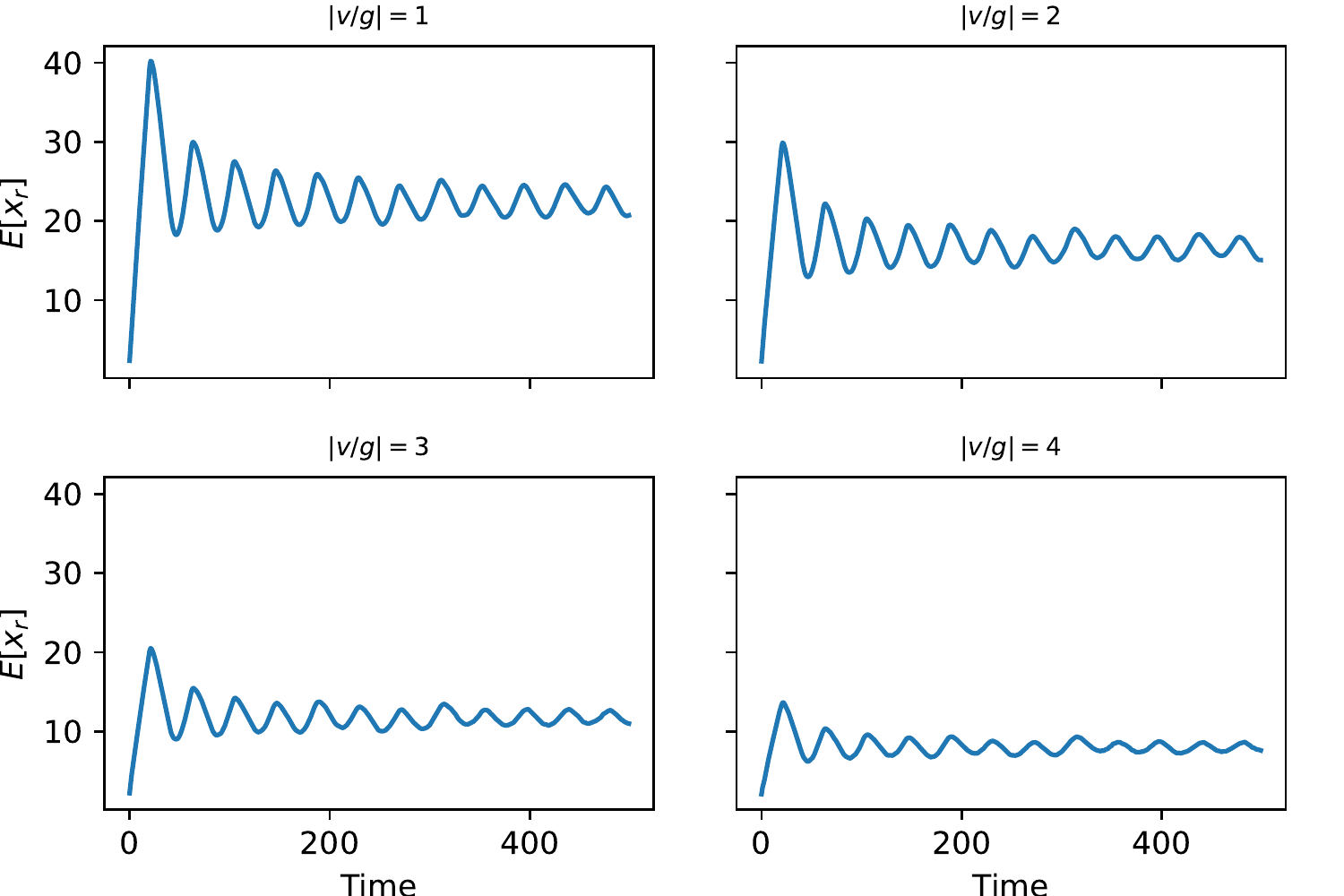}
        \caption{Simulation of triplet state electron pair dynamics using exact diagonalization on a 1-D lattice with 81 sites starting one site apart at the center for different values of $v/g$, the ratio of Coulomb strength to hopping strength. $E[x_r]$ is the expected value of the relative coordinate averaged over the common coordinate. Larger values of $v/g$ bind the pair more strongly.}
    \label{meanrelpos}
\end{figure}

For a large enough ratio $v/g$, the two electrons can never fully escape one another and delocalize. Consider the electron wavefunction expressed in relative and common coordinates, $x_r$ and $x_c$. The Hamiltonian in eq. (\ref{Coulomb}) is separable in this basis, so we can focus on the relative coordinate. We define localization to mean that there is a region of $x_r$ around $x_{r,0}$ where there is a nonzero probability to find the electrons for all time, and delocalization when this is not the case. For electrons starting $x_{r,0} \neq 0$ sites away in $D$ dimensions, the initial potential energy is $v/x_{r,0}$, and the the initial kinetic energy $\langle T \rangle$ is zero. In order for the pair to fully delocalize, it would have to convert $v/x_{r,0}$ of potential energy to kinetic energy. This is because a delocalized wavefunction has all of its probability located arbitrarily far from the origin where the Coulomb potential is zero \footnote{The fact that we required the onsite potential to be finite is actually necessary for this argument to work.$\bra{k}V\ket{k} \neq 0$ if $V$ is a true divergent Coulomb potential because the infinite potential at the origin cancels out the vanishing normalization factor of the delocalized $k$ state. In other words, the Fourier transform of $1/r$ is not zero.}. The maximum allowed kinetic energy on the lattice is $2Dg$ per electron, so if $v>4x_{r,0}Dg$ then the pair will remain localized forever. The critical starting distance for a localized pair is therefore $x_{r,0} = v/(4Dg)$. Note that this logic is completely equivalent to figuring out when an electron would have \emph{enough} energy to escape a positron. Simulations of bound triplet state eletronium in 1-D are show in Figs. \ref{meanrelpos} and \ref{satdistance} for different Coulomb strengths and starting positions, and they show agreement with the ideas presented in this section. Increasing their Coulomb repulsion and reducing their starting distance both cause the electrons bind more strongly.

\comments{
\begin{figure}
    \includegraphics[scale = .565]{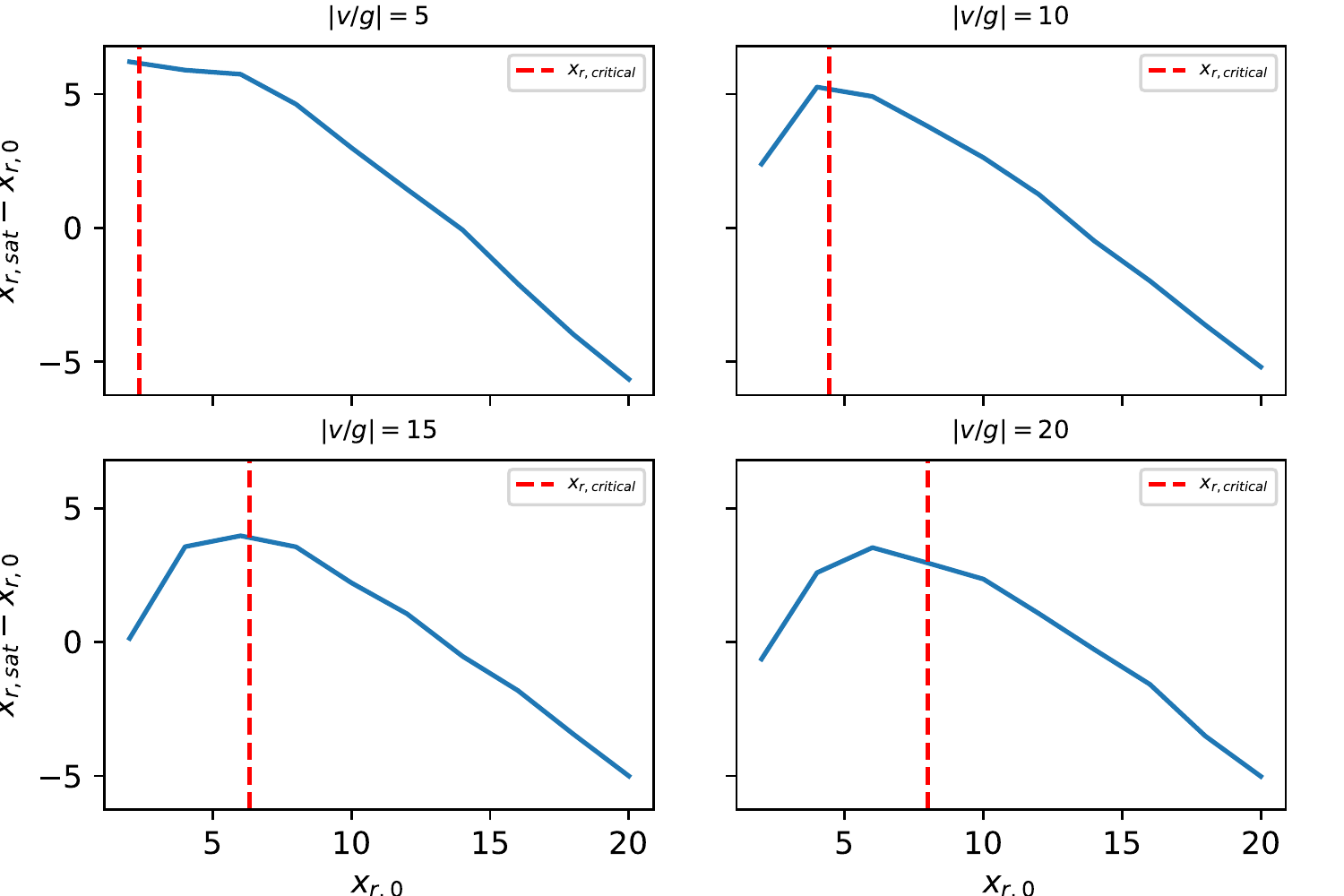}
    \caption{Simulation of triplet state electron pair asymptotics using exact diagonalization for a 1-D lattice with 81 sites and different values of $v/g$. Electrons are initialized on a single site a distance $x_{r,0}$ away with the same phase and time evolved according to the Hamiltonian in eq. (\ref{Coulomb}). The saturation distance $x_{r,sat}$ is $E[x_r]$ averaged over long times $t=499$, $t=699$. Red vertical lines are $x_{r,critical} = (2g/v+1/40)^{-1}$, the maximum starting distance where energy conservation says the pair would fail to delocalize past $x_r=40$. These plots show that smaller starting distances below $x_{r,critical}$ cause smaller saturation distances. Beyond $x_{r,critical}$, finite size pressure reduces the slope of $dx_{r,sat}/dx_{r,0}$ below $1$, indicating that the wavefunction has delocalized enough to feel the walls of the lattice.}
    \label{satdistance}
\end{figure}
}

\begin{figure}
    \includegraphics[scale = .58]{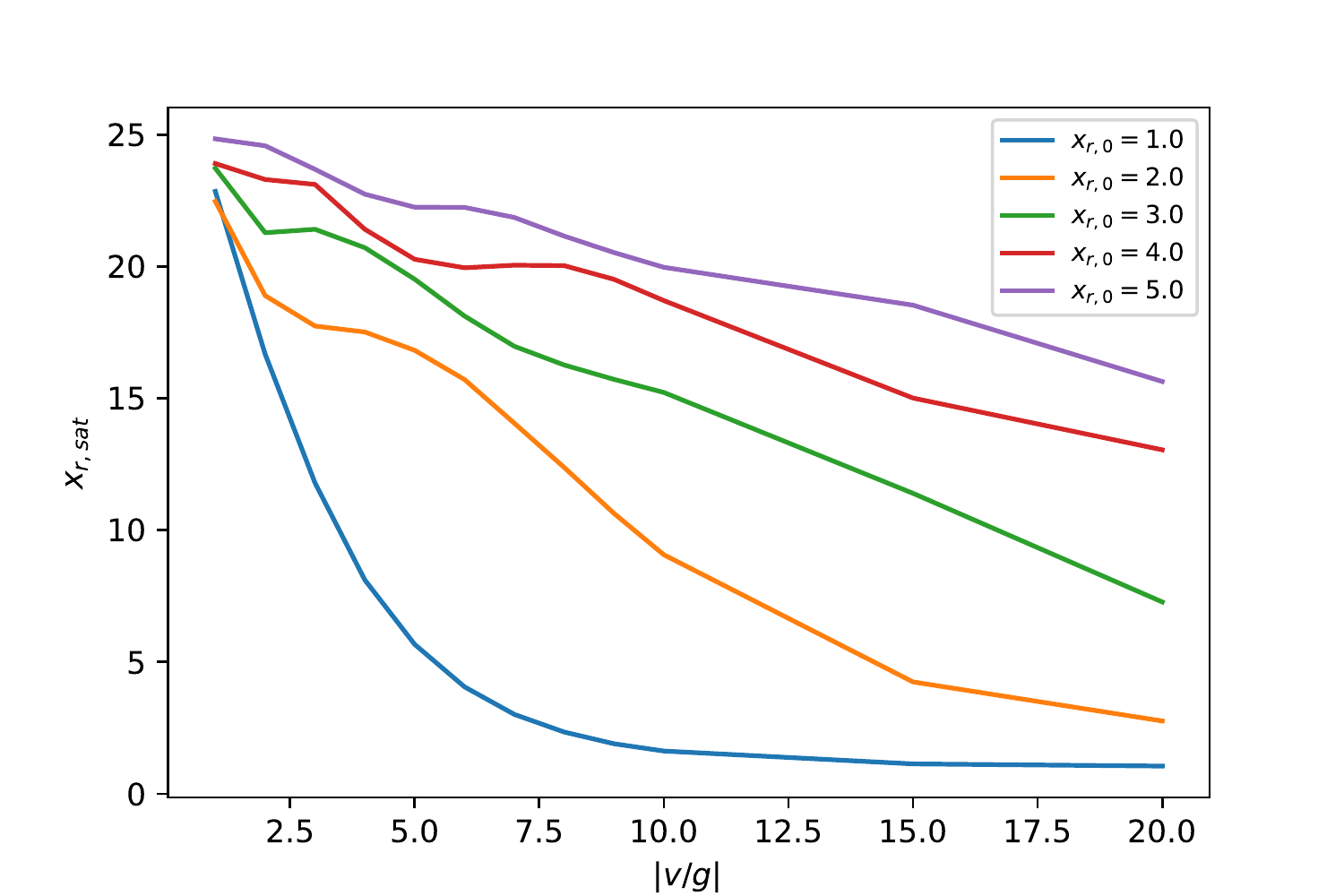}
    \caption{Simulation of triplet state electron pair saturation distance $x_{r,sat}$ vs Coulomb strength $v/g$. Electrons are initialized on sites a distance $x_{r,0}$ away with no phase and time evolved according to the Hamiltonian in eq. (\ref{Coulomb}). The saturation distance $x_{r,sat}$ is $E[x_r]$ averaged over 200 times from $t=1300$ to $t=1500$. }
    \label{satdistance}
\end{figure}

If the tight binding hopping term of the Hamiltonian is meant to describe the continuum Schrodinger equation kinetic energy term $\nabla^2$, then the hopping strength $g$ will scale with the lattice spacing $a$ like $g = g_0/a^2$, and the potential strength $v$ will scale like $v_0/a$. This means that the critical distance for bound electrons will be $x_{r,0} = v_0a/(4Dg_0)$. 

\comments{When $x_{r,0} < \frac{vs}{2Dg}$, we know that the pair cannot fully delocalize, but we can be more specific and bound the mass of the wavefunction which is allowed to leave a given large radius $R$. In order for a fraction $|\phi_{loc}|^2 \equiv \int_{B_R(0)}dX_r \hspace{.05cm} |\phi(X_r)|^2$ of the mass to leave the ball of radius $R$ centered at the origin $B_R(0)$, the energy of the remaining mass $|\phi_{loc}|^2$ must go up. The minimum change in energy of the delocalized mass is given by $\Delta E_{deloc} < (\frac{v}{R}-\frac{v}{x_{r,0}}+2g) < 0$, which can always be made less than zero by choosing a large enough $R$. The upper bound on the energy the localized mass can gain is $\Delta E_{loc} < (v_{on}-\frac{v}{x_{r,0}}+2g)|\phi_{loc}|^2$ for a singlet state pair. In order to conserve energy, we need $\Delta E_{loc} + \Delta E_{deloc} = 0$, which can only happen if 

\begin{equation} 
|\psi_{loc}|^2 > (\frac{v}{x_{r,0}}-\frac{v}{R}-2g)/(v_{on}-\frac{v}{R})
\end{equation}

By taking the limit as $R$ goes to infinity, we find 

\begin{equation}
    |\psi_{loc}|^2 > (\frac{v}{x_{r,0}}-2g)/v_{on}
    \label{loc}
\end{equation}

This gives the minimum mass of the wavefunction which is square integrable on an infinite lattice as $t \rightarrow \infty$. A larger $v_{on}$ allows more of the wavefunction to delocalize by ``pushing" on the localized part, increasing its energy. For any finite $v_{on}$, there will always be a non-zero localized mass. Because of the scaling $v \propto \frac{1}{a}$ and $g \propto \frac{1}{a^2}$, the fraction of wavefunction which must remain localized grows with the lattice spacing, and approaches $1$ as $s \rightarrow \infty$. Electrons in a triplet state cannot access $v_ons$, so eq. ($\ref{loc})$ should be modified by replacing $v_{ons} \rightarrow 2$. 

}

In a typical solid, lattice spacings are around $a = $5\AA, so a discrete version of the continuum kinetic energy term would have strength $g = \frac{\hbar^2}{2m_e s^2} \approx .15$eV \footnote{This is not exactly what sets the nearest-neighbor interaction strength in solids, but it gives approximately the right energy scale}, while the Coulomb potential two lattice sites away would be $1eV$. In 3D we have $2Dg = .9$eV, so we can conclude that a Coulomb potential will localize two electrons that are initially two lattice sites apart. In real materials this is made more complicated by the fact that Coulomb potential is screened, and that electrons can dump energy by scattering. In order to observe this electron pair trapping, one would need to find a material with a small hopping strength/screening and a large coherence time. Alternatively, one could look for this effect in systems of interacting trapped ions, which are likely to have both long coherence times as well as small hopping strengths. 

\section{Quantum Field Theory}

The potential inversion theorem is useful in contexts where particles can be created or destroyed. In the spirit of the end of section \ref{PIT}, we can consider particle-number of a given species to be a discrete lattice, where creation and annihilation operators are nearest-neighbor hoppings. Specifically, for a Yukawa interaction between fermion field $\psi$ and scalar field $\Phi$ with $H_{int} = g\Bar{\psi} \phi \psi$, take $A = (-1)^{N_\Phi}$ with $N_\phi\equiv \int d^4 k(2\pi)^{-4} a^{\dagger}_k a_k.$ We have $A \psi = \psi A$, $A \Phi = -\Phi A$, and $A \Pi_\Phi = -\Pi_{\Phi} A$. Since $H_0 = H_{Klein-Gordon}$ contains only terms with an even number of $\Phi$ and $\Pi_\Phi$, we have $H_+ A \equiv (H_0+H_{int})A = A(H_0-H_{int}) \equiv AH_-$. The potential inversion theorem tells us that for in/out states with definite numbers of $\Phi$ particles before, $N_{\phi,i}$, and after, $N_{\phi, f}$, scattering matrix elements under $H_+$ and $H_-$ are related by $S_{if}^+ = S_{if}^- (-1)^{N_{\Phi,i}-N_{\Phi,f}}$. Therefore, $|S_{if}^+|^2 = |S_{if}^-|^2$, so the transition probability does not depend on the sign of the interaction. Corollary of this is the well-known result \cite{peskin2018introduction} that Yukawa theory is attractive independent of the sign of $g$. This can be seen by noting that $\Bar{\psi}\psi \rightarrow \Bar{\psi}\psi$ diagrams must contain an even number of interaction vertices. However, the potential inversion theorem has the advantage of making this symmetry manifest in a non-perturbative way. 

This procedure works in general for any Hamiltonian which contains only terms with even numbers of field operator of a given species, which for the right choice of species includes $\Phi^4$ theory, QED, QCD, BCS theory, the Hubbard model, the Heisenberg model, and many more. We can add an odd term $gH_{odd}$, and conclude that probability evolution is preserved under the transformation $g\rightarrow -g$ for in-states with definite numbers of particles. If $|\psi(f,t,g)|^2 \equiv |\bra{f} e^{-i(H_0 + gH)} \ket{i}|^2$ is analytic in $g$ around $g=0$, then we can conclude from its evenness that $|\psi(f,t,g)|^2 = |\psi(f,t,0)|^2 + \mathcal{O}(g^2)$ for a perturbatively small $g$.

\section{Spin Models}

Spin 1/2 models are a canonical way to study quantum phases of matter like magnetism and antiferromagnetism, and numerical studies of spin models are used to explore fundamental quantum statistical phenomena like many body localization and thermalization. These models can be mapped onto lattice hopping models, and vice-versa, where we can take advantage of the potential inversion theorem. Consider the transverse field Ising Hamiltonian for $N$ spins in one dimension:
\begin{equation}
    H =  \sum^N_{j = 1} -J\sigma_j^z \sigma_{j+1}^z - h \sigma^x_j
    \label{Ising}
\end{equation}



\comments{
Consider the basis states of the full system where each spin is either up or down in the z direction. Each of these states can be thought of as a binary string where the jth bit represents whether the jth spin is up or down.  We can now place these binary strings on a hypercube where each vertex represent one binary string and each edge of the cube represents flipping one spin up or down. This embedding of binary strings of length N into a N dimensional hypercube is called Hamming space.  Since the set of points in Hamming space represent an ortho-normal basis for the Hilbert space of the spin system, Hilbert space for the spin system and Hilbert space of single quantum particle propagating in Hamming space are isomorphic.

The parity of a binary string is 1 if that string contains an even number of ones and -1 otherwise. Hamming space is a bipartite lattice, in the sense that for a vertex with parity $1$ (even), all nearest neighbor vertices will have parity $-1$ (odd), and vice versa. In section \ref{gendim}, we define the $A$ operator in $N$ dimensional space as diagonal in position with value $A = (-1)^{\Sigma_i x_i} = \bigotimes \sigma_z$. If we define the binary string with all zeros to be located at the origin, this A operator simply multiplies each point in Hamming space by the parity of its binary string.  Moving along the edge of the hypercube to the nearest neighbor vertex corresponds to flipping a single spin/bit, and the specific spin which is flipped corresponds to which dimension was traveled along to get to the nearest vertex. This collection of vertices is therefore a lattice of length 2 and dimension $N$ on which the parity inversion theorem applies. Note that if we instead consider this same $A$ operator in the $x$ basis, it corresponds to flipping every spin since .

}
We can consider each combination of spins to be a corner of an N-dimensional hypercube as shown in Fig. \ref{spins}. Importantly, this mapping works even if the spin lattice is frustrated, like in the three spin example at the bottom of Fig. \ref{spins}. In the basis $z\equiv \{\uparrow_i^z, \downarrow_i^z\}$, $\sigma^x_i$ is a nearest-neighbor hopping, and $\sigma^z_i$ is a zero-length hopping. Specifically, we can use $A = (-1)^{N_z}$ with $N_z \equiv \sum_{j = 1}^N \sigma^z_j$, and we have $A\sigma_i^x = -\sigma_i^x A$ and $A \sigma_i^z = \sigma_i^z A$. The potential inversion theorem tells us that if $\psi_0$ is a superposition of states with strictly even or odd numbers of $\uparrow_z$, then $|\psi(z,t)|^2$ is conserved by the transformation $h\rightarrow -h$. It also tells us that the eigenstates transform under $h \rightarrow -h$ by multiplication with $A$, and their energies stay the same. If the initial conditions are real up to a global phase, then $|\psi(z,t)|^2$ is conserved under $J \rightarrow -J$. Under this transformation eigenstates are again transformed by $A$, but their energies are inverted. This is a well-known result that the high energy ferromagnetic states correspond to the low energy ferromagnetic states. If $h$ is treated as a small perturbation, then by the logic in the previous section $|\psi(z,t)|^2$ differs from the unperturbed probability to leading order by $O(h^2)$.

\begin{figure}
    \includegraphics[scale = .44]{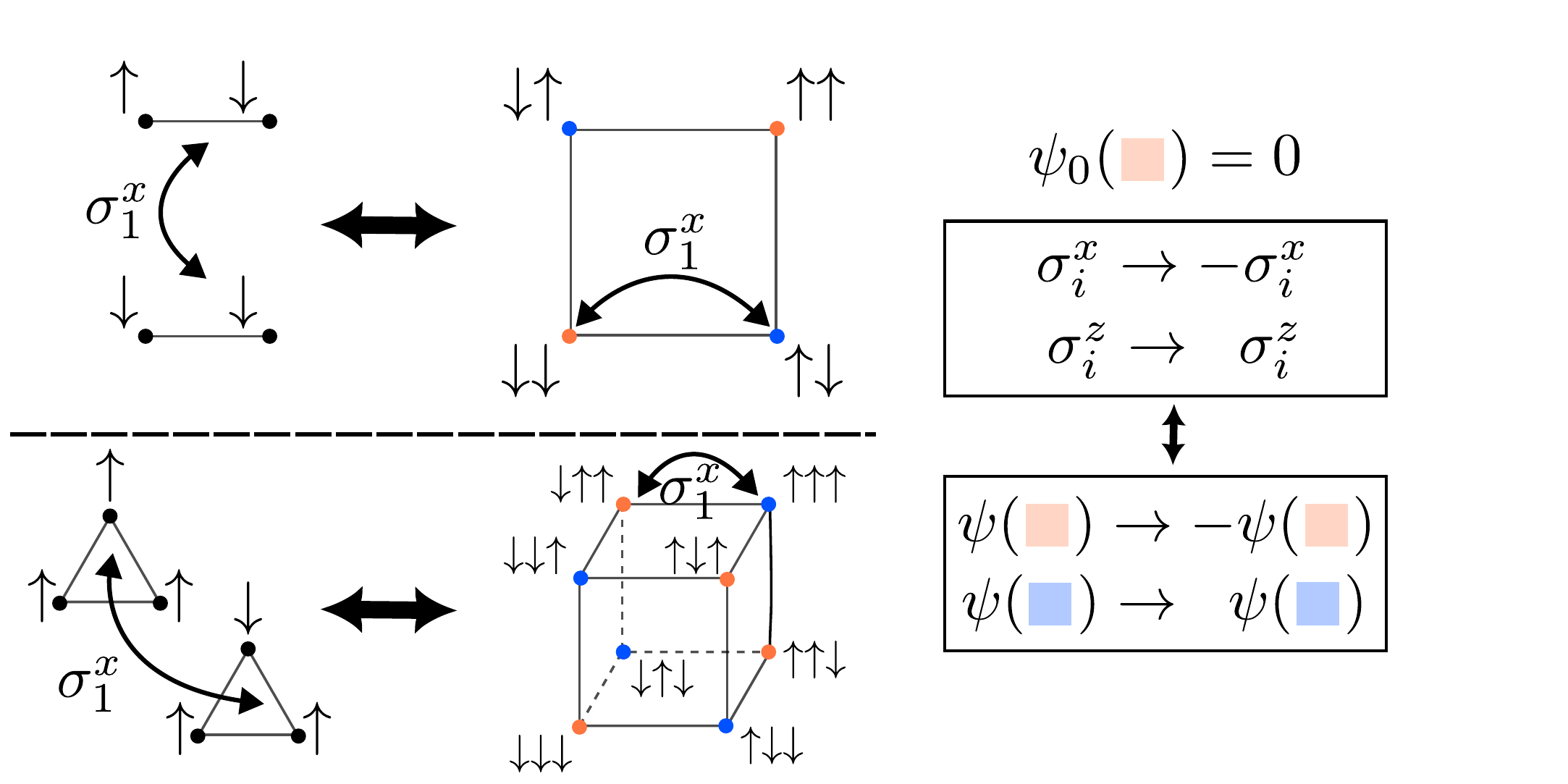}
    \caption{Representation of spin $1/2$ systems as vertices of a hypercube. In the top left, we show how $\sigma_1^x$ acts on a state with two spins, and on the right we see the same interaction represented as a nearest neighbor hopping on the square. On the bottom, we see the same except with three spins. Colors indicate the value of $A = (-1)^{N_z}$, and as in Fig. \ref{checkerboard}, the potential inversion theorem says that for initial conditions only on blue squares, inverting odd-distance hoppings inverts the value of the wavefunction on orange squares.}
    \label{spins}
\end{figure}

Alternatively, if we use $A = (-1)^{N_x/2}$, then $\sigma_i^x$ becomes a zero-distance hopping and $\sigma_i^z\sigma_{i+1}^z$ becomes a nearest-neighbor hopping. The potential inversion theorem then says that for initial conditions with $N_x/2$ strictly even or odd, $|\psi(x,t)|^2$ is conserved under $J \rightarrow -J$. If the initial conditions were real up to a global phase, then $|\psi(x,t)|^2$ is conserved under $h \rightarrow -h$ as well. Note that these conditions are different than the ones in the previous paragraph. If $J$ is treated as a small perturbation, then $|\psi(x,t)|^2$ differs from its unperturbed value to leading order by $O(J^2)$.

The potential inversion theorem can be used for any spin-interaction Hamiltonian made up of general N-body interactions between spins on the lattice, independent of interaction distance or lattice geometry. This additionally works for spins greater than $1/2$ if the Hamiltonian is built out of raising and lowering operators instead of Pauli matrices. Different choices of $A$ will map between the eigenvectors of a large class of Hamiltonians with the same spectra. For the most general Hamiltonian of $N$ interacting qubits, there are $3\times 2^N$ such choices of $A$. 

We can use these $A$ operators to map thermodynamic quantities of interest as well. Suppose $AH_+ = H_-A$, then we have $Z^+ = Tr(e^{-\beta H^+}) = Z^-$, and in a thermal state $\rho_{+} = e^{-\beta H^+}/Z = A\rho_- A$. This means that thermal expectation values of operators $O$ are related by $\langle O \rangle ^+ = \langle AOA \rangle $. For the Hamiltonian in eq. (\ref{Ising}), this corresponds to the statement that the expected magnetization $\sum_i \sigma_i^z$ for a ferromagnetic model $J>0$ at positive temperature $1/\beta$ is the same as the expected magnetization for the antiferromagnetic model with $-J$ at negative temperature $-1/\beta$.

\comments{

If we choose the spin $z$ basis for our system, then $-J\sigma_j^z \sigma_{j+1}^z$ is a zero-length coupling since it flips no spins, and $-h \sigma^x_j $ is a nearest neighbor coupling since it flips a single spin. Therefore the parity inversion theorem says that if $\psi_0$ is a sum of terms containing only even or odd numbers of up spins, then $|\psi(x, t)|^2$ is conserved by the transformation $h \rightarrow -h$, meaning the probability density of being in a certain $z$ basis configuration is identical for magnetic fields in opposite directions. If $\psi_0$ is real up to a global phase, then the corollary to the potential inversion theorem tells us that $|\psi(x, t)|^2$ is conserved by the transformation $J \rightarrow -J$ also, which implies configurations of spins have identical probability density with ferromagnetic and anti-ferromagnetic coupling.

\begin{figure}
    \includesvg[scale = .44]{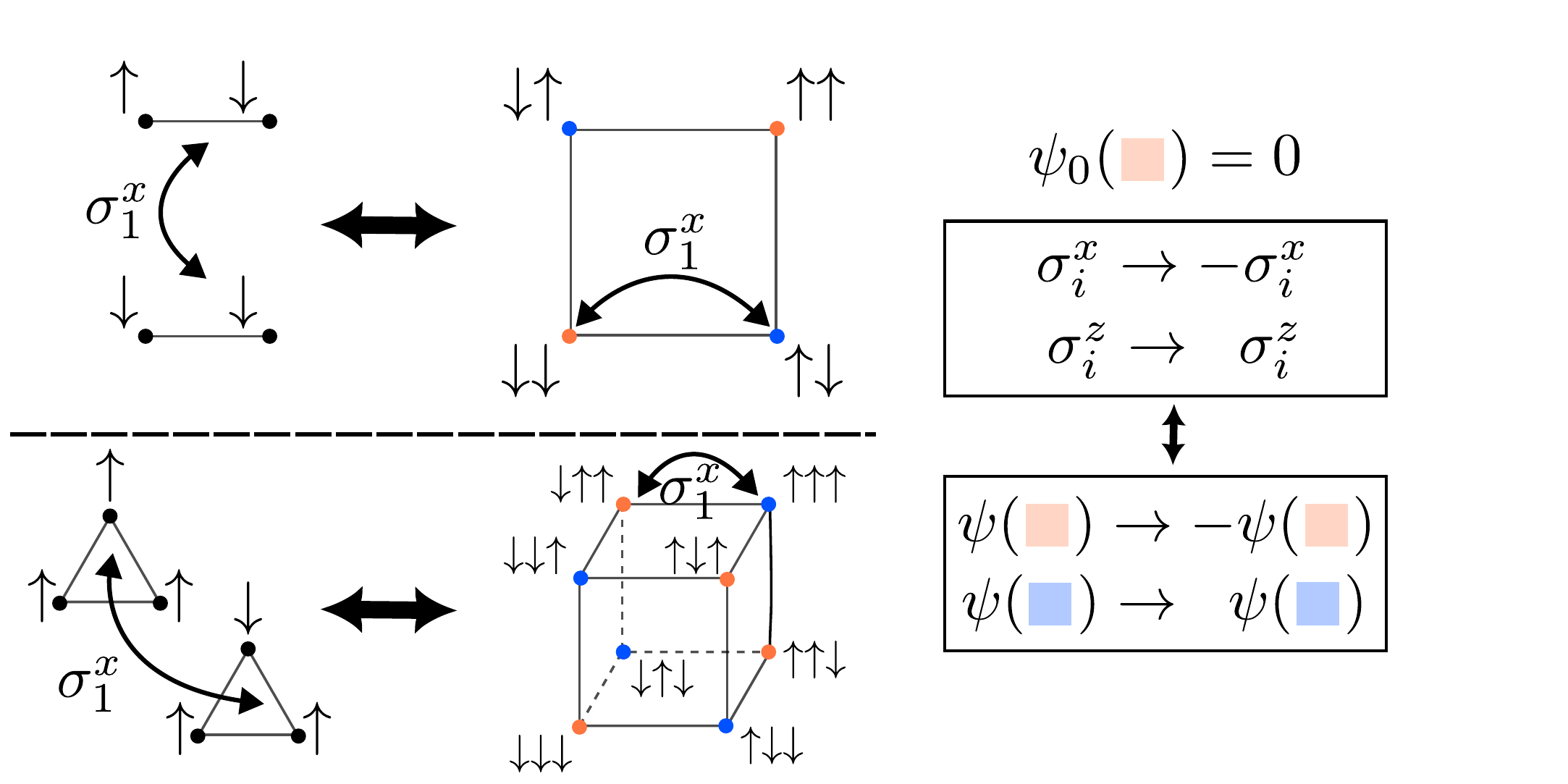}
    \caption{Representation of spin $1/2$ systems as vertices of a hypercube. In the top left, we show how $\sigma_1^x$ acts on a state with two spins, and on the right we see the same interaction represented as a nearest neighbor hopping on the square. On the bottom, we see the same except with three spins. Colors indicate the value of $A = (-1)^{N_z}$, and as in Fig. \ref{checkerboard}, the potential inversion theorem says that for initial conditions only on blue squares, inverting odd-distance hoppings inverts the value of the wavefunction on orange squares.}
    \label{spins}
\end{figure}

If we instead consider basis states where every spin in an eigenvector of $\sigma^x_j$, and in embed the spins in Hamming space based on their $\sigma^x_{j}$ eigenvalues, then $-J\sigma_j^z \sigma_{j+1}^z$ is a next-nearest neighbor coupling, and $-h \sigma^x_j $ is a zero-length coupling. By considering the invariant subspace of states with parity 1, we can define a new parity $p'_x \equiv (-1)^{\text{\# of up spins /2}}$. With this new parity and lattice, we see that $-J\sigma_j^z \sigma_{j+1}^z$ becomes a nearest-neighbor coupling, so the parity inversion theorem tells us that if $\psi_0$ is a sum of states with the same $p'$ parity, then $|\psi(x,t)|^2$ is conserved by the transformation $J \rightarrow -J$. Note that this is a different condition than the one in the previous paragraph, because $\psi_0$ is not required to be real up to a global phase.

We can use the parity inversion theorem for general types of spin interaction Hamiltonians. These Hamiltonians can include terms of the form 

\begin{equation}
T_i = g_i \bigotimes_{j = 1}^d \sigma^{m_{i_j}}_{i_j} 
\end{equation}

Where $m_{i_j} \in \{x,y,z\}$, and $\sigma_{i_j}$ acts on the $i_j$'th site and is the only term in the product to do so. In the spin $z$ basis, each term in the product with $m_{i_j} \neq z$ is a sum of nearest-neighbor couplings, and each term with $m_{i_j} = z$ is a a zero-length coupling. If we call $n_{i}$ the number of times that $m_{i_j} \neq z$ in the product, then we can define the parity of $T_i$ to be $p_i \equiv (-1)^{n_i}$. The parity inversion theorem says that if $\psi_0$ is a sum of states with the same parity, then $|\psi(t)|^2$ is conserved by the transformation $g_i \rightarrow -g_i$ for any $T_i$ whose parity is -1. In addition, if $\psi_0$ is real up to a global phase, then the corollary to the potential inversion theorem says that $|\psi(t)|^2$ is conserved by the transformation $g_i \rightarrow -g_i$ for all $T_i$ whose parity is 1.

We can use the result from the end of section \ref{parity} to strengthen this symmetry to include transformations of the form $g_{i,k} \rightarrow -g_{i,k}$ for operators $T_{i,k}$ which contain either $\sigma^y_{k}$ or $\sigma^x_{k}$. The corresponding initial conditions are that $\sigma_k^z \psi_0 = \pm \psi_0$. Alternatively, we can consider the set of spins $\{n_k\}$ and invert all operators $T_i$ which contain an odd number of terms of the form $\sigma^y_{\{n_k\}}$. The corresponding initial conditions are that $\bigotimes_j \sigma^z_{n_j} \psi_0 = \pm \psi_0$. The parity inversion theorem is agnostic to the geometry and dimension of the spin lattice, including interaction distance, and the only thing it takes into account is which spins that the interaction flips. 

We can generalize these mappings even further by considering arbritrary spin instead of spin $1/2$. A chain of $N$ spin $1$ particles, for example, can be mapped onto an $N$ dimensional hypercube of length 3. Terms in the corresponding spin interaction Hamiltonian will still have the same interpretation as odd or even parity couplings on this new larger hypercube, so we can use the inversion theorems as before. This allows us to map any bipartite lattice model in any dimension onto an interacting spin model, with the corresponding symmetries given by the inversion theorems.
}

\comments{
The fact that $H^+$ and $H^-$ are related by a simple mapping implies a simple mapping between their density matrices in the thermally disordered states. The density matrix for a thermally disorder state of the Hamiltonian $H$ is $\rho_{th} = e^{-\beta H}/Z$.  Now consider the density matrix for the thermally disorder state of $H^+$ as mapped through $A$.
\begin{align}
    A\rho_{th,+}A = \frac{1}{Z^+}Ae^{-\beta H^+}A \\
    = \frac{1}{Z^+}e^{-\beta AH^+A} \\
    = \frac{1}{Z^+}e^{-\beta H^-}
\end{align}
Now considering the partition function:
\begin{align}
    Z^+ = \Tr({e^{-\beta H^+}}) \\
    = \Tr(AA {e^{-\beta H^+}}) \\
    = \Tr (Ae^{-\beta H^-}A) \\
    = Z^-
\end{align}
Combining these results:
\begin{align}
    A\rho_{th,+}A = \frac{1}{Z^-}e^{-\beta H^-} = \rho_{th,-}
\end{align}
This implies that the entropy as a function of temperature only is identical for the two systems since A is unitary map:
\begin{align}
    S^+ = -\Tr(\rho_{th,+} \ln(\rho_{th,+})) \\ 
    = -\Tr(AA \rho_{th,+} AA \ln(\rho_{th,+})) \\
    = -\Tr(A \rho_{th,+} A \ln(A \rho_{th,+} A)) \\
    = -\Tr(\rho_{th,-}\ln(\rho_{th,-})) = S^-
\end{align}

This mapping also implies a simple relation between the expectation value of observables in the thermally disordered state.
\begin{align}
    \langle O \rangle^- = \Tr[O \rho_-] = \Tr[O A \rho_+ A] \\
   = \Tr[A O A \rho_+] = \langle A O A \rangle^+
\end{align}
The expectation values of observables in the disordered phase are also related by the $A$ transformation.

}

\section{Continuum Limit}

Outside of the negative scattering conditions shown in section \ref{Neg}, the potential inversion theorem fails in the continuum limit and shows that the continuous Schrodinger equation is a singular limit of discrete Schrodinger equations. In the discrete case, inverted potentials can trap particles just as well as positive potentials, as shown by the evolution of electron pairs shown in Figs. \ref{satdistance} and \ref{meanrelpos}. This clearly is not the case in the continuum limit.

When discretizing the Schrodinger equation $H = \partial_x^2 + V(x)$ with a lattice spacing $a$, we can use the finite difference formula to write $\partial_x^2$ as a nearest-neighbor hopping $T$. This gives $H = -T/a^2+V(xa)+1/{a^2} \equiv T'+V'$. For real initial conditions, the potential inversion theorem says that $|\psi(x,t)|^2$ is preserved by $V' \rightarrow -V'$, but in the continuum limit this corresponds to inverting the potential $V$ and subtracting a divergent energy. We can attempt to patch this problem by considering the Hamiltonian $\partial_x^2-\partial_y^2+V(x)$, where these divergent terms cancel. However, neither the substitution $V(x) \rightarrow -V(x)$ nor $\partial_x^2-\partial_y^2 \rightarrow -\partial_x^2+\partial_y^2$ preserve the evolution of $\psi_0(x) = \delta(x)$, despite both of these transformations being allowed in the discrete setting.  There is proveably no $A$ operator which commutes with $V(x)$ and anticommutes with $\nabla^2$ in the continuum. In addition, there are many initial conditions which time-evolve differently on a lattice but approach the same Dirac delta function in the limit of lattice spacing going to zero. For example, two Kronecker delta functions located next to each other vs two sites away. For this reason the intuition from the potential inversion theorem must break down in the continuum for any situation in which large $k$ components of the wavefunction play a major role (notably not the case in section \ref{Neg}).
 
A nearest-neighbor hopping Hamiltonian with a linear potential (such as that for an electron in a semiconductor with a voltage difference at the two ends) is known to exhibit Bloch oscillations, where the electrons oscillate back and fourth sinusoidally. This is in stark contrast to the expected Ohm's law behavior, and is an example of the potential inversion theorem. In Ohmic materials decoherence stops this process, as the electrons can scatter off of phonons to release their energy. In a sense, this means that resistive heating \emph{causes} current, and not the other way around. In the same way, two electrons trapped next to each other in a solid can only escape by getting rid of their energy with scattering. This is not the case in the continuum, where two electrons can increase their kinetic energy without bound as they escape. 

The potential inversion theorem can be used for discrete forms of other PDE's by converting derivatives to hoppings using finite difference formulas. For example, the heat equation $\partial_t u = \nabla^2 u$ would function identically to the Schrodinger equation on a lattice with regards to $A$, except instead of oscillating eigenstates we have decaying heat modes. Alternatively, we could add a non-Hermitian decay term $D = -\Gamma \sum_{x,\sigma} a^{\dagger}_{x,\sigma}a_{x+\Delta x}$ to the hopping Hamiltonian in eq. (\ref{one}), and we would have $AD = DA$ if $\Delta x$ was even, and $AD = -DA$ if $\Delta x$ was odd. We would then have $e^{-i(H_++D)t}\ket{\psi_0} = Ae^{-i{H_--D}t}\ket{\psi_0}$ as usual.

\section{Discussion}

The potential inversion theorem can be thought of as a consequence of conservation of energy. For a lattice hopping model, any potential which varies over the lattice by more than the allowed kinetic energy range can localize particles. In the same way that the frequency response of a damped oscillator is symmetric about the resonant frequency, too much energy prevents a wavefunction from coherently traversing a gap. In the context of Anderson localization, the potential inversion theorem explains why the high energy band as well as low energy band are localized. Inverting the potential conserves probability evolution, and inverting a disordered potential landscape with mean energy $0$ causes it to look the same statistically over a large distance. Therefore if the lowest energies are localized, then so are the highest. Our results show that the Hamiltonians which allow for delocalization are actually rare, because they not only have to have low enough disorder (small $V(k)$ for large $k$), but they also have to not varry significantly over large distances (small $V(k)$ for small $k$).

The $A$ operator which multiplies by $-1$ on every other site can be thought of as adding $\pi$ to the wave-vector $k$, taking $k = 0$ to the edge of the Brouillon zone, and $k = \pi$ back to the center. This turns the lowest kinetic energy wave-vectors into the highest, and vice versa. An electron trapped in $V(x) = x^2$ will have high kinetic energy with low potential energy in the center, and low kinetic energy with high potential energy near the edges. The wavefunction of the electron trapped in the inverse potential $V(x) = -x^2$ is just $A$ times the electron in the regular potential. This corresponds to the inverse potential electron having low kinetic energy with high potential energy in the center, and high kinetic energy with low potential energy at the sides, which is exactly what one would expect.

To simulate the continuous Schrodinger equation with a nearest neighbor hopping model, a small enough lattice size must be chosen so that wavevectors near the Brouillon zone boundary play an unimportant role. Otherwise the potential inversion theorem will lead to spurious results such as electrons getting trapped on the peaks of potential landscapes. The point where this error sets in is when the change in potential is larger than the allowed change in kinetic energy. Since the hopping strength $g$ should scale like $1/{a^2}$, this can be avoided by making the lattice smaller. For a large enough $g$, the kinetic energy $2-2gcos(k) \approx gk^2$, and the hopping energy will represent the continuum kinetic energy. The period and amplitude of Bloch oscillations both decrease with lattice spacing, so for a given time one can always find a small enough lattice spacing where the oscillations do not appear. The critical distance for binding of electron pairs on a lattice also goes down with momentum cutoff, and the mass of the electron wavefunction which is able to escape goes up (see the discussion at the end of section \ref{epbs}).

The potential inversion theorem fails on a frustrated lattice (like a triangular lattice). When there is no way to assign parity to lattice sites like in Fig. \ref{checkerboard}, we will not be able to find an operator $A$ which makes the theorem work. This is not just a roadblock to a proof, but is in fact representative of the fundamentally different physics of bipartite and frustrated lattices. However, interacting spins can always be mapped onto a non-frusterated lattice like in Fig. \ref{spins}, which could be useful for studying spin-glass models and other exotic phases.

We provided a dynamical mechanism for electrons and holes to form bound states on a lattice with only Coulomb repulsion in both singlet and triplet configurations. Bound states of this kind could potentially lead to superconductivity, and will be investigated further in future works.

\section*{Acknowledgements}
We thank Daniel Tartakovsky for his extensive mentorship and guidance over the course of writing this paper, as well as Michael Peskin for reviewing it and offering helpful and insightful suggestions. We would also like to thank George Crowley for thought provoking discussions.

\comments{
\section{Linear Potential}

In this section we will see how a linear potential in 1-D localizes eigenstates and traps excitations, as well as how this localization scales with system size, hopping strength, and potential strength. We find that localization occurs for potentials whose strength at the system boundary is more than ~5 times the hopping strength. We also find that the size of eigenstates scales linearly in the ratio of hopping strength to potential strength. We will show how this leads to excitations being trapped in a region which scales as the inverse lattice spacing cubed as the discrete Schrodinger equation approaches the continuous one. 

To start, we discuss why the eigenstates of $H_+ \equiv T+a\hat{x}$ will all be localized by a strong enough $a$. The eigenstates will clearly be localized in the positive $x$ direction since the potential energy gets too high, and the potential inversion theorem shows us why they will be localized in the $-x$ direction as well. This is because the eigenstates of $H_+$ can be mapped onto the eigenstates of $H_-$ by A, and the eigenstates of $H_-$ must decay in the negative direction due to the potential energy being too high. We find numerically that the size of the middle energy eigenstate is given by $\sigma_x = \sqrt{2}/a$ with $g$ fixed. Additional numerical results confirm that the eigenstate size scales linearly in $g$ while keeping $a$ fixed, and linearly in $g/a$ while varying both $a$ and $g$. 

\comments{
\begin{figure}
    \centering
    \includegraphics[scale = .55]{eigenstates3.png}
    \caption{Size of middle eigenstate for the linear potential with hopping strength $g = -1$ and various system sizes. $\sigma_x \equiv \sqrt{<x^2>}$ is the standard deviation of position measured in lattice sites. Adding more lattice sites doese not change the size of the eigenstate for a given $a$ since it is too localized to notice the edges of the system. The relation $\sigma_x = \sqrt{2}/a$} holds to nine decimal places.
    \label{eig3}
\end{figure}
}

Due to their localization, eigenstates near the center are approximately translations of one another. This is because with translation operator $D$ and Hamiltonian $H_+ \equiv T+a\hat{x}$, we have

\begin{equation}
    H_+ D \ket{\phi_n,+} \approx D(T+(a+1)\hat{x})\ket{\phi_n,+} = (E_{+,n}+1)\ket{\phi_{+,n}}
\end{equation}

The above equation is only approximately an equality because in a finite system the displacement operator destroys some of the state. However, the localized eigenstates near the center of the system have almost no value at the boundaries, so this approximation is extremely accurate. The degree to which the eigenstates near the center are translations of one another will be used as a simple test of localization, and will predict whether an initial excitation will remain localized. 

\begin{figure}
    \centering
    \includegraphics[scale = .55]{eigenstates2}
    \caption{Plot of all 1000 eigenvectors overlaid for $L=1000$, $g=-1$, and various values of a. The flat region in the middle corresponds to eigenvectors which are almost translations of one another. When this translation relationship breaks down due to edge effects, the peak of the modulus squared will change between nearest eigenvectors, causing the flat region to abruptly become curved. The size of this flat region quantifies how well the eigenstates are localized, since the edge of the flat region is the first eigenstate to extend far enough to feel edge effects. }
    \label{eig2}
\end{figure}

\begin{figure}
    \centering
    \includegraphics[scale = .55]{eigenstates1}
    \caption{Plot of all $L$ eigenvectors overlaid for $g = -1$, various values of $L$, and the critical potential strength $a = 10/L$ where localization sets in. Above this strength, the flat region becomes large enough to sustain the full periodic oscillation of an initial excitation at the origin, and below this strength, the excitation will get to a size where it notices edge effects through the non-translational eigenvectors. }
    \label{eig1}
\end{figure}

In Fig. \ref{eig2}, we have plotted every eigenstate for $L=1000, g=-1,$ and different values of $a$. We can clearly see a flat region in the center of the graph corresponding to eigenstates being approximately translations of one another. This region grows with $a$ as the eigenstates become more strongly localized. The critical value of $a$ which will cause localization of an initial excitation at the center of the lattice is given by $10/L$, corresponding to a value of $ax = 5$ at the boundary of the lattice. Figure \ref{eig1} demonstrates that the size of the flat region only depends on this maximum value of $ax$ and not on $L$ itself.

Intuitively, an excitation at the center of the lattice cannot diffuse because it is comprised only of sums of these localized, translational eigenstates. It does not see any eigenstates which could carry it to the edge, unlike the case for small $a$. The critical value of $a = 10/L$ for localization corresponds to a flat eigenstate region about the same size as the saturation length for an excitation diffusing for small $a$. This transition from localized to delocalized excitations is shown in Fig. \ref{diffusion1}. In this Fig., an initial excitation at the central lattice site is time evolved, and the time-average of its standard deviation is plotted for several values of $a$.

\begin{figure}
    \centering
    \includegraphics[scale = .55]{diffusion2}
    \caption{Diffusion of an initial excitation located at the central lattice site for $g = -1$ and various values of $a$. x is normalized to lattice length, i.e. $-1/2,x,1/2$ so that $\sigma_x$ can be compared for different values of $L$. We see that for values of the potential strength below the critical $a = L/10$ allow the initial state to expand to the point where aperiodic decoherence occurs, causing a failure to relocalize at the center of the lattice. Above the critical strength, the diffusion is periodic and relocalizations occur for all time (tested up to $t = 10**9$).}
    \label{diffusion2}
\end{figure}

\begin{figure}
    \centering
    \includegraphics[scale = .55]{diffusion1}
    \caption{Periodic diffusion of an initial excitation located at the central lattice site for $g = -1$ and various values of $a$. $x$ is normalized to lattice length, i.e. $-1/2,x,1/2$. We see that for values of the potential strength below the critical $a = L/10$ allow the initial state to expand to the point where aperiodic decoherence occurs, causing a failure to relocalize at the center of the lattice. Above the critical strength, the diffusion is periodic and relocalizations occur for all time (tested up to $t = 10**9$).}
    \label{diffusion1}
\end{figure}

In Fig. \ref{diffusion1}, we can see the dynamics of the diffusion. Delocalized states experience decoherence from interactions with with the boundary, and they fail to relocalize at the center. The time evolution of the localized states is periodic, and corresponds to real-space Bloch oscillations.  The time evolution of $\sigma_x^2$ is shown in Fig. \ref{diffusion3} and is well fit by $-A\cos{(\frac{ft}{a})}+A$ with $A = 4g^2/a^2L^2$ and $f = a$. Since $\sqrt{A} = 2g/La$ is the amplitude of the $\sigma_x$ oscillations, we have the same scaling of frequency and amplitude with potential strength as we would have with Bloch oscillations. The Bloch oscillation of an electron in a system with lattice spacing $s$ and electric field $E$ can be written 

\begin{equation}
    x(t) = x_0 + \frac{2g}{eE}\cos{(\frac{seEt}{\hbar})}
\end{equation}

Whereas the oscillations in our model are 

\begin{equation}
    \sigma_x(t) =   \frac{2g}{La}|\sin{(\frac{at}{\hbar})}|
\end{equation}

These equations are analogous, with $La = a/s$ corresponding to $eE$, the potential energy per lattice site distance.

\begin{figure}
    \centering
    \includegraphics[scale = .55]{diffusion3}
    \caption{Periodic diffusion of an initial excitation located at the central lattice site for $g = -1$ and various values of $a$. $x$ is normalized to lattice length, i.e. $-1/2,x,1/2$. These plots show that diffusion for a strong enough potential can be thought of as a position space Bloch oscillation with amplitude of $\sigma_x$ given by $\sqrt{A} = 2/aL$ and frequency $f = a$.}
    \label{diffusion3}
\end{figure}
}

\comments{
\section{Introduction (Version 2)}
The evolution of the Schrodinger equation on discrete lattices is a topic of wide interest. In the tight binding models of solids, an electron can be in a superposition of the atomic orbitals of each atom in the lattice.  To simplify, we imagine that each atom in the lattice only contains one atomic orbital, which either filled or empty and that electrons can hop between neighbouring atoms. Any vector in the Hilbert space for a single electron in the tight binding lattice can then be expressed as vector with a complex entry for every site in the lattice where its magnitude represents the square root of probability that the electron is on that atom and the phase of the complex number represents the phase of the Schrodinger wave.  Here we demonstrate a surprising result that the if one considers any potential applied to a 1D lattice, the probability density of the electron being on given lattice site is identical regardless of whether you invert the potential or not for two initial conditions related by a simple diagonal matrix.  In addition, we show that the eigen-energies of the original and inverted Hamiltonians are simply the opposite of one another and the eigenvectors are related by a simple diagonal matrix.
\section{Inverse Potential Theorem for States (Version 2)}
\subsection{Theorem}
Consider the Schrodinger equations

\begin{equation}
    \begin{cases}
            i\partial_t \psi_{\pm}(x,t) = H_{\pm} \psi(x,t)
 & t \in (-\infty,\infty)\\
             \psi_{\pm}(x,0) = \psi_{\pm,0}(x) &  x \in \mathbb{N}, 1\leq x \leq L \\
             \psi_{+,0}(x) = (-1)^x\psi_{-,0}(x)
    \end{cases}
    \label{SE11}
\end{equation}

Where $\psi_{\pm}(t) \in \mathbb{C}^L$ and $H_{\pm}: \mathbb{C}^L \rightarrow \mathbb{C}^L$ are $L\times L$ Hermitian matrices given by 

\begin{equation}
H_{\pm} = T \pm V
\end{equation} 

Where $T$ is the Hermitian nearest neighbor hopping matrix with $g$ straddling the diagonal

\begin{equation}
    T^r_c = g\delta^r_{c+1}+g\delta^r_{c-1}
\end{equation}

and $V$ is any Hermitian potential energy matrix which is diagonal in the position basis, i.e.

\begin{equation}
V^r_c = v(r) \delta^r_c
\end{equation}

for some $v: \{ x | x\in \mathbb{N}, 1\leq x \leq L \} \rightarrow \mathbb{R}$. Then $\forall t\in (-\infty,\infty)$, we have 

\begin{equation}
    |\psi_+(x,t)|^2 = |\psi_-(x,-t)|^2
\end{equation}
and more precisely

\subsection{Proof}
The central object in this proof is the diagonal matrix:
\begin{equation}
    A^r_c = (-1)^r \delta^r_c
    \label{Aeqn1}
\end{equation}
\begin{align}
    \psi_+(x,t) = (-1)^{x} \psi_-(x,-t)
\end{align}
The matrix $A$ has two important properties: it commutes with any $V(x)$ since its diagonal in the position basis, and it anti-commutes with $T$. Define the anti-commutator 
\begin{equation}
    \{A,T\}^r_c \equiv (AT+TA)^r_c = 0
\end{equation}
Equivalently,
\begin{equation}
    TA = -AT
\end{equation}
This can be shown as follows:
\begin{equation}
\{A,T\}^r_c = g[(-1)^c(\delta^r_{c+1}+\delta^r_{c-1})+(-1)^r(\delta^r_{c+1}+\delta^r_{c-1})] = 0
\end{equation}
This means that $A$ can be used as a diagonal change of basis matrix to express $H_+$ in terms of $H_-$:
\begin{equation}
    H_{+} A = TA + VA = -AT+AV = -A(T-V) = -AH_{-}
\end{equation}
Since A is idempotent
\begin{equation}
    -A^{-1}H_+ A = -A H_+ A = H_-
\end{equation}

The solution of eq. (\ref{SE1}) is 
\begin{equation}
    \psi_{\pm}(x,t) = \bra{x} e^{-iH_{\pm}t} \ket{\psi_0} \equiv \bra{x} U_{\pm}(t) \ket{\psi_0}
\end{equation}
Conjugating the evolution operator by the change of basis A:
\begin{equation}
    AU(t,0)A = A e^{-iH_{+}t} A = e^{iH_{-}t} = U(-t,0)
\end{equation}
This implies
\begin{equation}
    U(t,0)A = A U(-t,0)
\end{equation}
Now looking at the evolution of the initial condition $\psi_{0,+}$ and using the relation.
\begin{equation}
    \psi_{0,+}(x) = (-1)^x \psi_{0,-}(x) = \sum_{x'} A^x_{x'} \psi_{0,-}(x')
\end{equation}
So, the evolution due to $H_+$ can be expressed:
\begin{align}
    \psi_+(x, t) = \sum_{x'} U_+(t,0)^x_{x'} \psi_{0,+}(x) \\
    = \sum_{x', x''} U_+(t,0)^x_{x'} A^{x'}_{x''} \psi_{0,-}(x'') \\
    = \sum_{x', x''} A^{x}_{x'} U_-(-t,0)^{x'}_{x''} \psi_{0,-}(x'') \\
    = \sum_{x'} A^{x}_{x'} \psi_{-}(-t)(x') = (-1)^x \psi_{-}(x,-t)
\end{align}
Therefore,
\begin{align}
    \psi_+(x,t) = (-1)^{x} \psi_-(x,-t)
    \label{wave_rel}
\end{align}
taking the absolute value squared of this expression:
\begin{align}
    |\psi_+(x,t)|^2 = |\psi_-(x,-t)|^2
\end{align}
\section{Statement of Theorem}

Consider the discrete Schrodinger equations on L sites

\begin{equation}
    \begin{cases}
            i\partial_t \psi_{\pm}(x,t) = H_{\pm} \psi(x,t)
 & t \in (-\infty,\infty)\\
             \psi_{\pm}(x,0) = \psi_{\pm,0}(x) &  x \in \mathbb{N}, 1\leq x \leq L \\
             \psi_{+,0}(x) = (-1)^x\psi_{-,0}^*(x)
    \end{cases}
    \label{SE11}
\end{equation}

where $\psi_{\pm}(t) \in \mathbb{C}^L$ and $H_{\pm}: \mathbb{C}^L \rightarrow \mathbb{C}^L$ are $L\times L$ Hermitian matrices given by 

\begin{equation}
H_{\pm} = T \pm V
\end{equation} 

Where $T$ is given by

\begin{equation}
    \bra{x} T \ket {y} \equiv g\delta^y_{x+1}+g\delta^y_{x-1}
\end{equation}

for some $g \in \mathbb{R}$ and $V$ is any diagonal real matrix. Then $\forall t\in (-\infty,\infty)$, we have 

\begin{equation}
    |\psi_+(x,t)|^2 = |\psi_-(x,-t)|^2
\end{equation}

\section{Proof}

Consider the diagonal matrix $A$ given by 

\begin{equation}
    \bra{x} A \ket{y} \equiv (-1)^x \delta^y_x
\end{equation}

$A$ commutes with $V$ because both are diagonal, and $A$ anticommutes with $T$ because 

\begin{equation}
    \bra{x}\{A,T\}\ket{y} \equiv \bra{x}(AT+TA)\ket{y}
\end{equation}

\begin{equation}
= g[(-1)^x(\delta^y_{x+1}+\delta^y_{x-1})+(-1)^y(\delta^y_{x+1}+\delta^y_{x-1})] = 0
\end{equation}

This means that 

\begin{equation}
    AH_+ \equiv A(T+V)
\end{equation}

\begin{equation}
    = (-T+V)A = -H_-A
\end{equation}

Consequently, the time evolution operator $U_+ \equiv e^{-iH_+t}$ can be related to $U_-$ via

\begin{equation}
    e^{-iH_+t}A \equiv \sum^\infty_{n=0} \frac{(-iH_+t)^n}{n!}A
\end{equation}

\begin{equation}
     =A\sum^\infty_{n=0} \frac{(iH_-t)^n}{n!} = A e^{iH_-t}
\end{equation}

The solution of eq. (\ref{SE11}) is the time-evolved initial condition

\begin{equation}
    \psi_{+}(x,t) = \bra{x}e^{-i H_+ t}\ket{\psi_{+,0}}
    \label{56}
\end{equation}

Because $\psi_{+,0}(x) = (-1)^x\psi_{-,0}^*(x)$, we have $\ket{\psi_{+,0}} = (A \ket{\psi_{-,0}})^* $, so eq. (\ref{56}) becomes

\begin{equation}
  = \bra{x}e^{-i H_+ t}A\ket{\psi_{-,0}}^* = \bra{x}Ae^{i H_- t}\ket{\psi_{-,0}}^*
\end{equation}

Since $A$ is hermitian, we can act it to the left

\begin{equation}
=(-1)^x \bra{x} e^{iH_-t} \ket{\psi_{-,0}}^* = 
\end{equation}
}

\clearpage
\bibliography{sorsamp}
\end{document}